\documentclass[useAMS,usenatbib,a4paper]{mn2e}

\usepackage{amsfonts,amssymb,amsmath}
\usepackage{aas_macros}
\usepackage{times,varioref,color,multirow,textcomp}
\usepackage[pdftex]{graphicx}
\usepackage[
    pdftex,% get better results and can break links but can't use psfrag
    a4paper=true,
    plainpages=true,
    pdfpagelabels,
    breaklinks=true,
    bookmarks=true,
    bookmarksopen=false,
    bookmarksopenlevel=2
    bookmarksnumbered=true,
    bookmarkstype=toc,
    colorlinks=true,
    citecolor=red,
    linkcolor=blue,
    menucolor=green,
    urlcolor=magenta,
]{hyperref}

\hypersetup{
  pdfauthor = {Pascal Jahan Elahi},
  pdfkeywords = {Code, Numerical},
  pdftitle = {STructure Finder}
}

\def \Msun{\ {\rm M_\odot}}
\def \R{\mathcal{R}}
\def \ELL{\mathcal{L}}
\def \Nse{N_{\rm se}}
\def \Nv{N_{\rm v}}
\def \Ncell{N_{\rm cell}}
\def \ELLth{\ELL_{\rm th}}
\def \Vr{\mathcal{V}_{\rm r}}
\def \Thetaop{\Theta_{\rm op}}
\def \ellx{\ell_{\rm x}}
\def \ellv{\ell_{\rm v}}
\def \erf{{\rm erf}}
\def \logphase{\langle\log[F({\bf X})/f({\bf X})^{\rm NN}_{\rm bg}]\rangle}
\def \logrho{\langle\log[\rho({\bf x})/\rho({\bf x})^{\rm NN}_{\rm bg}]\rangle}
\def \volfrac{\mathcal{V}_f}
\def \veliso{v_{\rm iso}}
\def \STF{{\sc stf}}

\def \FOF{{\sc fof}}
\def \6DFOF{{\sc 6dfof}}
\def \enlink{{\sc enlink}}
\def \hsf{{\sc hsf}}
\def \Gadget2{{\sc gadget-2}}
\def \galactics{{\sc galactics}}
\newcommand{\Eqref}[1]{Eq.~(\ref{#1})}
\newcommand{\Figref}[1]{Fig.~\ref{#1}}
\newcommand{\Secref}[1]{\S\ref{#1}}  
\newcommand{\Tableref}[1]{Table~\ref{#1}}

\defcitealias{nfw}{NFW}
\defcitealias{enbid}{{\sc enbid}}
\defcitealias{skid}{{\sc skid}}
\defcitealias{enlink}{{\sc enlink}}
\defcitealias{subfind}{{\sc subfind}}
\defcitealias{hsf}{{\sc hsf}}

\begin{document}
\title[STructure Finder]{Peaks above the Maxwellian Sea: A New Approach to Finding Substructure in N-Body Haloes}
\author[P.J.~Elahi, R.J.~Thacker, and L.M.~Widrow] {Pascal~J.~Elahi$^1$, Robert~J.~Thacker$^2$, Lawrence~M.~Widrow$^3$\\
$^1$Shanghai Astronomical Observatory, Chinese Academy of Sciences, Shanghai, China; \href{mailto:pelahi@shao.ac.cn}{pelahi@shao.ac.cn}\\
$^2$Department of Astronomy \& Physics, Saint Mary's University, Halifax, Nova Scotia, Canada;\\
$^3$Department of Physics, Engineering Physics \& Astronomy, Queen's University, Kingston, Ontario, Canada;
}
\maketitle

\begin{abstract}
We describe a new algorithm for finding substructures within dark matter haloes from N-body simulations. The algorithm relies upon the fact that dynamically distinct substructures in a halo will have a {\em local} velocity distribution that differs significantly from the mean, {\em i.e.} smooth background halo. We characterize the large-scale mean field using a coarsely grained cell-based approach, while a kernel smoothing process is used to determined the local velocity distribution. Comparing the ratio of these two estimates allows us to identify particles which are strongly cluster in velocity space relative to the background and thus resident in substructure. From this population of outliers, groups are identified using a Friends-of-Friends-like approach. False positives are rejected using Poisson noise arguments. This approach does not require a search of the full phase-space structure of a halo, a non-trivial task, and is thus computationally advantageous. We apply our algorithm to several test cases and show that it identifies not only subhaloes, bound overdensities in phase-space, but can recover tidal streams with a high purity. Our method can even find streams which do not appear significantly overdense in either physical or phase-space.
\end{abstract}
\begin{keywords}
methods: data analysis -- methods: numerical -- galaxies: haloes -- galaxies: structure -- dark matter
\end{keywords}
\maketitle

%---------------------------------
\section{Introduction}\label{sec:intro}
The dark matter in the haloes that surround galaxies such as our own is often modeled as a smooth distribution in both position and velocity space. In particular, the spatial distribution is generally assumed to fall off monotonically from the halo center, and the local velocity distribution is assumed to be Maxwellian. These assumptions provide a useful starting point for making model predictions for dark matter detection experiments ({\em e.g.}~\citealp{lewin1996}). However, the haloes that form in cosmological simulations are anything but smooth, even relaxed, virialized haloes that have not undergone any major mergers in recent history. In particular, they contain subhaloes, which range in mass from a few percent of the host halo's mass down to the resolution limit of the simulation ({\em e.g.}~ \citealp{moore1999,gao2004,diemand2006,diemand2008,springel2008}).

\par
Subhaloes are but one example of substructure found in simulated haloes. As subhaloes traverse the host they will experience mass loss and leave behind debris in the form of shells and streams. In simulations these coherent streams of particles can exist for a few orbital periods before eventually dissolving into the smooth halo background. The resulting superposition of streams and subhaloes produces a complex velocity structure (e.g.~\citealp{fairbairn2009,kuhlen2010}), and even the ``smooth'' background of a halo is composed of many overlapping fundamental dark matter streams \citep{stiff2003,vogelsberger2010}. Lastly, and perhaps most practically, the number, velocity profile and mass of these streams will have important ramifications for direct and indirect dark matter searches ({\em e.g.}~\citealp{stiff2001,vogelsberger2008,vogelsberger2009,kuhlen2010}).

\par 
The interest in streams is not limited solely to dark matter. Galactic stellar haloes are thought to form, in part, from the accretion and subsequent disruption of satellite galaxies \citep{johnston2008}. A prime example is the Sagittarius stream associated with the Sagittarius dwarf \citep{ibata1994,ibata2001}. The PANDAS survey has found evidence of stellar streams in M31, which may come from disrupted satellites \citep{pandas2009}.  These stellar substructures offer an observable window into galaxy formation history and searches for such structures are just beginning in earnest. Data from future instruments such as GAIA \citep{gaia2008} will allow us to test the $\Lambda$CDM model by comparing the number of observed stellar streams to the number predicted from cosmological simulations.

\par
The problem with finding substructures is non-trivial and a number of techniques have been developed to find subhaloes. These include {\sc skid} \citep{skid}, {\sc subfind} \citep{subfind}, \6DFOF\ \citep{diemand2006}, \enlink\ \citep{enlink}, and \hsf\ \citep{hsf}. Algorithms such as \citetalias{subfind} search for physical overdensities or clustering in configuration space and are well suited to finding subhaloes. Simulations show that the contrast between substructures and the background is more prominent in phase-space and action space \citep{helmi2000,gomez2010}. Algorithms such as \citetalias{enlink} and \citetalias{hsf} make use of this increased contrast by using phase-space density to identify substructures. In general, most of these algorithms search for local peaks in either physical or phase-space density and estimate the outer boundaries enclosing these peaks. A number of studies have searched for completely disrupted subhaloes by identifying particles that were once bound to a subhalo at some early time and that are no longer bound to this subhalo \citep{helmi2003,vogelsberger2009b,xu2009}. This process is laborious and computationally intensive since past snapshots must be stored and tracked from output to output.

\par
Here we present a new algorithm, the STructure Finder (\STF) designed to find velocity substructures within dark matter haloes. Substructures occupy specific regions in orbital space and have smaller velocity dispersion than the virialized background and consequently should appear more strongly clustered in velocity space relative to the background. In short, we identify candidate streams (and subhaloes) by finding groups of particles that "stand out" against the velocity distribution of the background. We model the local velocity distribution as a multivariate Gaussian and therefore, our method is most sensitive to substructures on the tails of the Gaussian, that is, substructures that are cold but that have a large bulk velocity. Once we have identified these outlying particles we then use a Friends-of-Friends-like algorithm to link them.

\par
Our paper is organized as follows: In section \Secref{sec:algorithm}, we outline our algorithm, determine the algorithm's optimal parameters, and discuss what types of substructure can be identified. We test our algorithm with toy models and discuss the results in \Secref{sec:toymodels}. In section \Secref{sec:halomany}, we analyze a more realistic model of a dark matter halo. The paper concludes in \Secref{sec:discussion} with a summary and discussion.

\section{The STructure Finder Algorithm}\label{sec:algorithm}
The STructure Finder (\STF) is designed to find velocity substructures in N-body haloes by identifying regions of the velocity distribution that differ from the smooth ``Maxwellian'' background. Since subhaloes are remnants of isolated haloes that formed at an earlier epoch, they should be physically denser and have smaller velocity dispersions than the host in which they now reside, and therefore have higher phase-space densities. In principle, if we could accurately measure the phase-space structure of a halo, both subhaloes and tidal streams would be easy to identify since Liouville's theorem states that their phase-space volume and density remains constant as the phase-space region they occupy is distorted by gravitational evolution. 

\par
Unfortunately, current N-body simulations do not have the resolution required to measure phase-space structure at the accuracy needed to identify diffuse tidal streams\footnotemark \citep{stiff2001}, though they can easily resolve the presense of subhalos \citep{knebe2011}. Furthermore, the highly anisotropic nature of the velocity distribution of streams makes it difficult to accurately measure their phase-space density. Finally, from a practical perspective, the techniques used to estimate the local phase-space density are substantially more computationally intensive than those used to estimate the physical density \citep{fiestas,enbid,enlink}.
\footnotetext{Technically, the volume is only conserved for the fine-grained distribution function and N-body simulations sample the coarse-grained distribution function. Furthermore, N-body codes do not explicitly solve the collisionless Boltzmann equantion and thereby explictly conserve this volume.}

\par
Our scheme relies on the assumption that a halo's velocity distribution can be split into a virialized background and substructures. Specifically, we assume that within a small volume of a simulated halo the distribution function is approximately separable so that one can write
\begin{align}
  F({\bf x},{\bf v})=\rho({\bf x})f({\bf v}),
\end{align}
where $\rho({\bf x})$ and $({\bf v})$ are the physical and velocity density, respectively. The ratio of the distribution function of particles associated to a substructure with a dispersion $\sigma_{\rm S}$ to that of the background distribution with $\sigma_{\rm bg}$ at the same phase-space coordinates is 
\begin{align} 
  \frac{F_{\rm S}({\bf x},{\bf v})}{F_{\rm bg}({\bf x},{\bf v})}=\frac{\rho_{\rm S}({\bf x})}{\rho_{\rm bg}({\bf x})}\frac{\sigma_{\rm bg}^3}{\sigma_{\rm S}^3}\frac{e^{-({\bf v}-{\bf v}_{\rm S})^2/2\sigma^2_{\rm S}}}{e^{-({\bf v}-{\bf v}_{\rm bg})^2/2\sigma^2_{\rm bg}}},
\end{align}
where here we have assumed the velocity distributions of both the substructure and the background are Gaussian. 

\par
This ratio has three distinct terms, the physical density contrast, the contrast in the dispersions and a ratio of Gaussian terms. Subhaloes are physically dense and dynamically cold. Conversely, tidal streams may have a negligible density contrasts and may also have velocity dispersion comparable with the background. In both cases, the substructure's velocity distributions will differ from the background. Consequently, although the Gaussian term for the substructure will likely be of order unity, due to the offset in the velocity, $\delta {\bf v}={\bf v}-{\bf v}_{\rm bg}$, this ratio is enhanced an factor of $\exp(\delta v^2/2\sigma_{\rm bg}^2)$. It is this exponential factor that is key to our algorithm. 

\par
Thus instead of measuring the phase-space density we measure the {\em local} velocity density of a particle, $f_{\rm l}({\bf v})$, and divide out the expect velocity density of the background at the particle's velocity, $f_{\rm bg}({\bf v})$. This ratio is used to identify particles that belong to velocity distributions which differ from the background. This ratio is particularly effective at identifying substructures with large relative bulk velocities regardless of their actual physical or phase-space density.

\par
\begin{figure}
    \centering
    \includegraphics[width=0.5\textwidth]{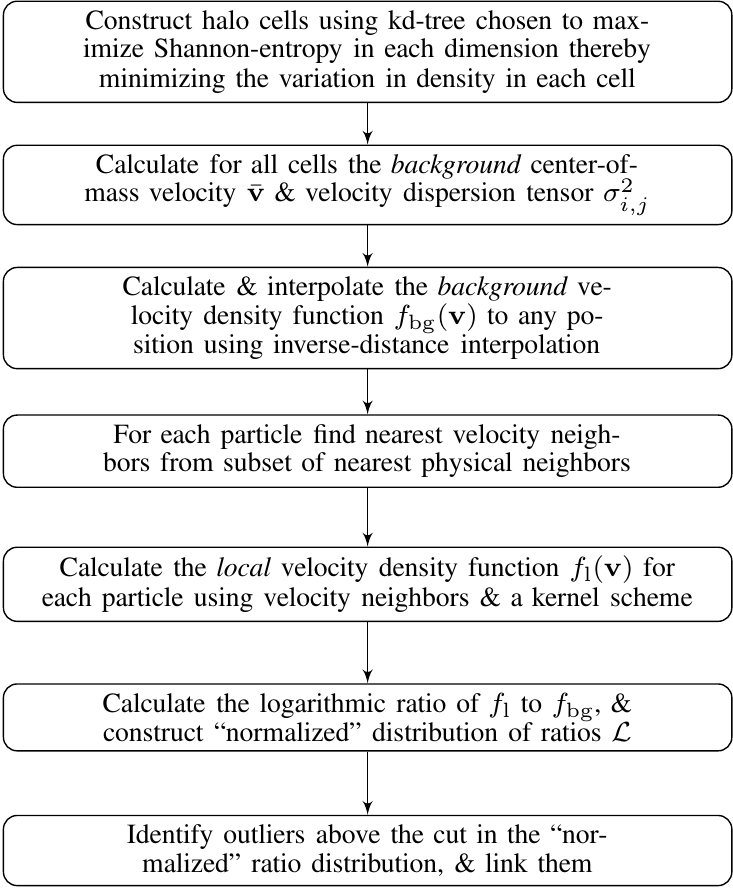}
    \caption{Flow chart of \STF\ algorithm.}
    \label{fig:stfflowchart}
\end{figure}

The \STF\ algorithm is composed of two main parts: the first part determines the likelihood of a particle lying outside the background velocity distribution; the second links these outliers using a Friends-Of-Friends algorithm. We summarize the steps of our algorithm in \Figref{fig:stfflowchart}. 

\subsection{Finding Outliers}\label{sec:outliers}
\subsubsection{Background velocity distribution}
We assume the mean background velocity density is characterized by a Multivariate Gaussian\footnotemark. Thus the expected {\em background} velocity density for a particle $k$ with velocity ${\bf v}_k$ is
\begin{align}
    f_{\rm bg}({\bf v}_k)=\frac{\exp\left[-\tfrac{1}{2}({\bf v}_k-\bar{\bf v})\Sigma^{-1}({\bf v}_k-\bar{\bf v})\right]}{(2\pi)^{3/2}|\Sigma|^{1/2}},\label{eqn:fvbg}
\end{align}
where $\bar{\bf v}$ is the average velocity, and $\Sigma$ is matrix representation of the velocity dispersion tensor about $\bar{\bf v}$, $\sigma^2_{i,j}$. Both $\bar{\bf v}$ and $\sigma^2_{i,j}$ depend of position relative to the halo's center-of-mass. 
\footnotetext{\cite{vogelsberger2009b} found that the velocity distribution from the central region of a high resolution simulation of a galactic halo was approximately characterized by a Multivariate Gaussian. The observed deviations away from this prediction may be due to the presence of substructure and may depend on the details of the halo's formation history. Hence a reasonable first order assumption is that the background is described by a Multivariate Gaussian.}

\par
The key to accurately characterizing the mean field is splitting the halo into appropriately sized volumes or cells. These volumes should contain enough particles so that the statistical error within a cell is negligible, but not be so large that density and potential varies greatly across the cell. The requirement that cells not be too large can be understood as follows: In a perfectly spherical, isotropic halo, $\bar{\bf v}=0$ but $\sigma^2\simeq GM(r)/r$. If the halo is decomposed into spherical shells and the dispersion calculated using all the particles in a shell, the estimator for $\sigma^2$ is effectively an average of $GM(r)/r$ over the radial thickness of the shell. Using a very thick shell will result in a biased estimator.

\par
To ensure that a balance is achieved between these competing effects, we split the halo into cells using a kd-tree \citep{kdtree,appel1985,treecode} so that each cell contains $\Ncell=N/2^a$ particles, where $N$ is the number of particles in the halo and $a$ is the depth of the tree. The tree is constructed by iteratively splitting the system along the spatial dimension that maximizes Shannon entropy, $S$. This quantity is estimated for each physical dimension in a volume containing $n$ particles by binning them in $n_{\rm bins}$ that span the extent of the dimension using the formula
\begin{align}
    S=\frac{1}{\log n_{\rm bins}}\sum\limits_k^{n_{\rm bins}}-\frac{m_k}{m_{\rm tot}}\log\frac{m_k}{m_{\rm tot}},
\end{align}
where $m_k$ is the mass in the $k^{th}$ bin and $m_{\rm tot}$ is the total mass. From the perspective of information theory, this process amounts to splitting the system along the dimension that contains the maximum amount of information. In this case, the splitting dimension will be the one with the greatest amount of variation in the particles' positions. This method effectively minimizes the variation in particle density across any given cell volume.

\par
Once the halo has been split into cells, we calculate the center-of-mass velocity and the velocity dispersion tensor for each cell,
\begin{align}
    \bar{\bf v}&=\frac{1}{M_{\rm cell}}\sum\limits_{k=1}^{\Ncell} m_k{\bf v}_k,\label{eqn:meanvel}\\%\quad 
    \sigma^2_{i,j}&=\frac{1}{M_{\rm cell}}\sum\limits_{k=1}^{\Ncell} m_k\left(v_{k,i}-\bar{v}_i\right)\left(v_{k,j}-\bar{v}_j\right)\label{eqn:dispvel},
\end{align}
where $M_{\rm cell}$ is the mass contain in the cell. 

\par
Estimating $f_{\rm bg}({\bf v}_k)$ using only the $\bar{\bf v}$ and $\sigma^2_{i,j}$ of the cell containing the particle can give rise to grid effects, that is sharp transitions in $f_{\rm bg}$ at a given velocity between cells. These discontinuities are  particularly noticeable for cells dominated by a single substructure. To minimize these effects, we interpolate $\bar{\bf v}$ and $\sigma^2_{i,j}$ (technically we interpolate the inverse of the velocity dispersion tensor, $\Sigma^{-1}$) at a particle's physical position with a inverse-distance interpolation scheme using the cell containing the particle and the six nearest neighbouring cells since the mean field should vary smoothly from cell to cell. Explicitly, we use 
\begin{align}
    u(\mathbf{x}) = \sum_{i = 0}^{N}{ \frac{ w_i(\mathbf{x}) u_i } { \sum_{j = 0}^{N}{ w_j(\mathbf{x}) } } },
\end{align}
where $u$ is the quantity we wish to determine at a position $\bf x$ based on cells with center-of-mass positions ${\bf x}_i$, and $w_i(\mathbf{x}) = |\mathbf{x}-\mathbf{x}_i|^{-1}$. 

\subsubsection{Local velocity distribution}
We estimate the {\em local} velocity density of a particle, $f_{\rm l}({\bf v}_k)$, using a kernel-scheme with an Epanechnikov smoothing kernel \citep{enbid}. This density is calculated using $\Nv$ nearest velocity neighbours from the set of $\Nse$ nearest physical neighbours, where $\Nv\leq\Nse$. So as to not introduce any grid effects we do not limit the search for a particle's nearby physical neighbours to particles that are in the same cell. Using a small number of velocity neighbours from a larger set of physical neighbours can give a biased estimate of the local velocity density. However, since our goal is to highlight clustering in velocity space, this is perfectly acceptable. We measure the local velocity density using a small number of nearest velocity neighbours, ie: $\Nv\sim10-100$. The set of physical neighbours from which the local velocity density is estimated should be substantially larger than the number used to calculate the velocity density. We generally fix $\Nse=32\Nv$.

\par
In summary, this part of the algorithm has two key parameters, $\Ncell$ and $\Nv$, which effectively define the volumes used to measure the mean and local velocity densities.

\subsubsection{Comparing the distributions}
\begin{figure}
    \centering
    \includegraphics[width=0.45\textwidth]{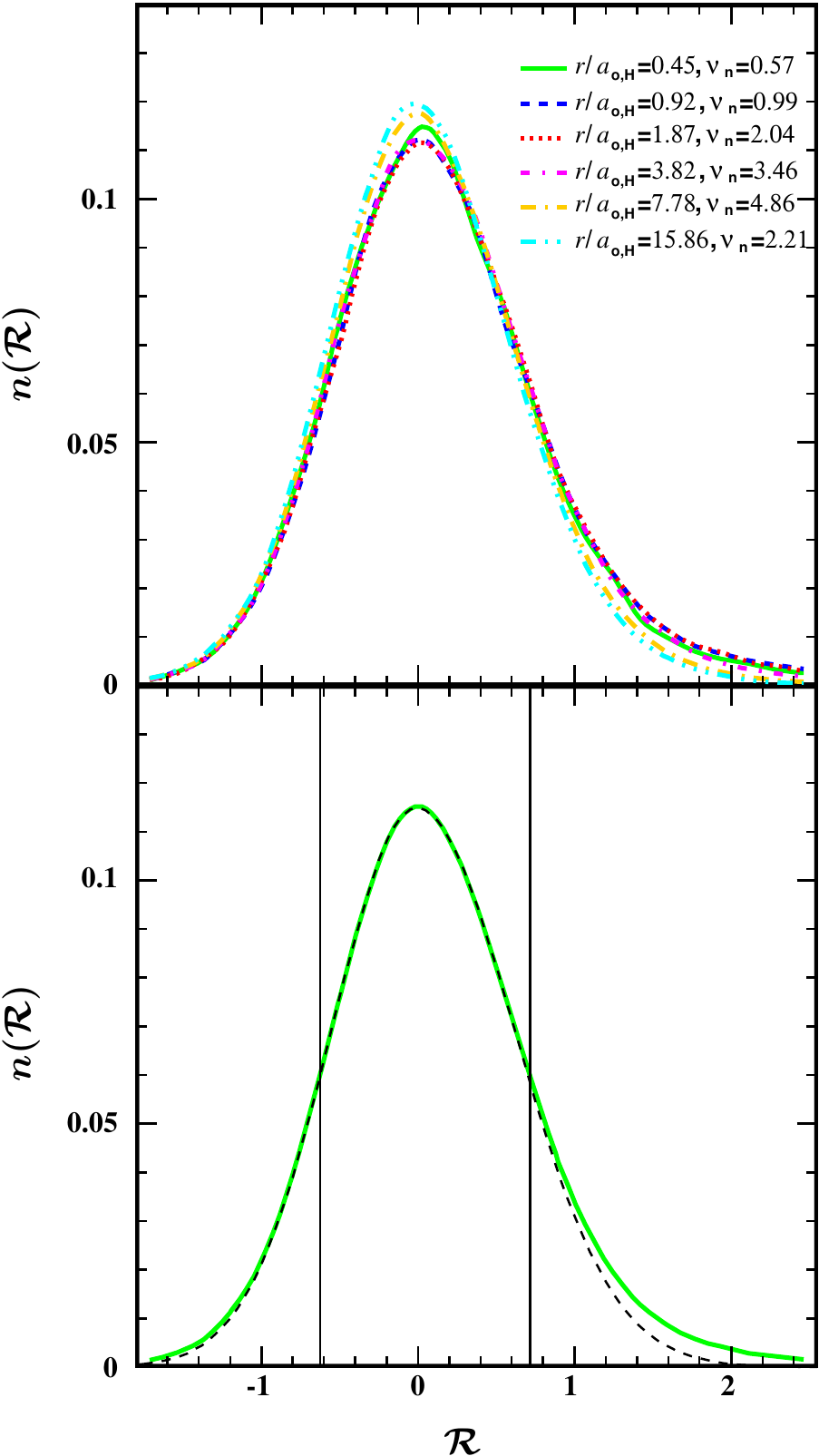}
    \caption{The normalized distribution of $\R$ for a smooth NFW halo. In the top panel wel plot the distribution in several different radial bins equally spaced in $\log r$ out to a radius of $1.4R_{\Delta}$, where $R_{\Delta}$ is the virial radius corresponding to an overdensity $\Delta=200$ and is at $11a_{o,{\rm H}}$. For each curve, we indicate the centre of the radial bin relative to the scale radius and also the number of particles in that bin $\nu_n$, in units of $10^3/{\rm kpc}^3$. In the bottom panel we show the global distribution (solid line), along with the Skewed-Gaussian fit (dashed line). Also shown are two vertical lines at approximately $\bar\R-s\sigma$ and $\bar\R+\sigma$ corresponding to the region that is used to determine the scatter of the background distribution, $\sigma_\R$.}
    \label{fig:Rresdistrib}
\end{figure}
Next, we consider the logarithmic ratio of the local and background velocity distributions,
\begin{align}
    \R_{k}=\ln\left[f_{\rm l}({\bf v}_k)/f_{\rm bg}({\bf v}_k)\right].
\end{align}
Before we address what $\R$-value indicates that a particle belongs to a substructure, we need to consider the scatter that may be present in a smooth halo without any substructure. 

\par
In order determine the form of $\R$-distribution for a smooth halo, we examine several smooth, spherical haloes generated by \galactics\ \citep{galactics1995,galactics,widrow2008}. \galactics\ generates self-consistent equilibrium models of spherical haloes with isotropic velocities and a density profile is given by
\begin{align}
    \tilde{\rho}_{\rm H}(r)=&\frac{2^{1-\alpha}V_o^2}{4\pi a_{o}^2} (r/a_{o})^{-\alpha}(1+r/a_{o})^{-(\beta-\alpha)}\notag\\ &\quad\times C(r,r_{\rm t}, \delta r_{\rm t}),\label{eqn:denprofile}
\end{align}
where $a_o$ is the characteristic radius of the halo, $V_o$ is the characteristic velocity dispersion, and $C$ is a truncation function that smoothly goes from unity to zero at the truncation radius $r_{\rm t}$ over a distance $\delta r_{\rm t}$. For an NFW \citep{nfw} density profile, $\alpha=1$ and $\beta=3$.

\par
In figure \ref{fig:Rresdistrib}, we show both the local and global $\R$-distribution for a galactic mass NFW halo composed of $10^6$ particles calculated $\Ncell=1953$ (that is 512 volumes), $\Nv=32$ ($\Nse=32\Nv=1024$). The $\R$-distribution does not appear to exhibit a strong radial dependence and is both locally and globally well characterized by a Gaussian, but is slightly skewed. However, it is important to recall that we use a biased estimator of the local velocity density which can introduce a skew or asymmetry in the distribution, even for a smooth halo. 

\par
For generality, we use a Skew-Gaussian to characterize the $\R$-distribution:
\begin{align}
    f_{\rm SG}&(\R;\bar{\R},\sigma_\R,s,A)=
    A\Biggl\{\exp\left[-\frac{\left(\R-\bar\R\right)^2}{2s^2\sigma_\R^2}\right]\Theta(\bar\R-\R)\notag\\&\quad+\exp\left[-\frac{\left(\R-\bar\R\right)^2}{2\sigma_\R^2}\right]\Theta(\R-\bar\R)\Biggr\},\label{eqn:skewgaus}
\end{align}
where $s$ is a measure of the skew or asymmetry, and $\Theta(x)$ is the Heaviside function. We fit this function to the binned distribution using a nonlinear least squares Levenberg-Marquardt algorithm, assuming the bins are independent and have Poisson errors. The fit shown in \Figref{fig:Rresdistrib} has a reduced $\chi^2\sim1$.

\par
\begin{figure}
    \centering
    \includegraphics[width=0.45\textwidth]{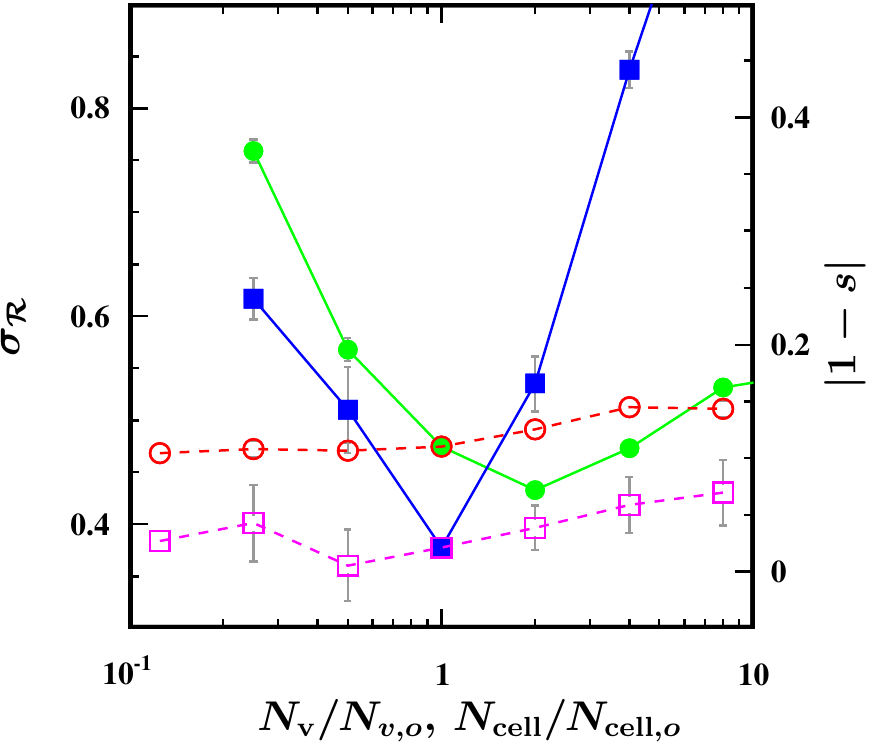}
    \caption{The variation in $\sigma_\R$ \& $s$ with $\Nv$ \& $\Ncell$ relative to fiducial parameters of $\Nv=32$ \& $\Ncell=1953$ (or 512 cells). Circles correspond to $\sigma_\R$, squares to $s$, filled symbols to $\Nv$ and open symbols to $\Ncell$.}
    \label{fig:Rresdistrib-param}
\end{figure}

Our primary goal at this point is to characterize, and ideally minimize, the dispersion in the background distribution, $\sigma_\R$. This parameter, along with the skew and mean, depends on both $\Nv$ and $\Ncell$. The optimal values for $\Nv$ and $\Ncell$ should minimize $\sigma_\R$ and $|1-s|$. In figure \ref{fig:Rresdistrib-param} we plot $\sigma_\R$ and $s$ as a function of $\Nv$ and $\Ncell$. This figure shows that the distribution is quite sensitive to $\Nv$, but significantly less so to $\Ncell$. At small $\Nv$, $\sigma_\R\propto\Nv^{-1/2}$, indicating that $\sigma_\R$ is dominated by the Poisson noise in the $f_l({\bf v})$ estimator. The optimal $\Nv$ appears to be 32-64. We find that this value appears to be independent of $\Ncell$, $\Nv/\Nse$, and the number of particles in the halo so long as $\Nse\lesssim1\%N_{\rm H}$ and $\Nse$ is a few times larger than $\Nv$. If one uses too large a value for $\Nse$, one will no longer be measuring the local velocity density function. The resulting $\R$ distribution will then be highly skewed. We have found that the local velocity density function begins to produce a strong bias or skew for $\Nse\lesssim256$. This implies that the minimum number of particles a halo must contain for our method to work effectively with little or no bias is $N_{\rm H}\gtrsim10^4$.

\par
The weak dependence of $\sigma_\R$ on $\Ncell$ indicates that $\sigma_\R$ is dominated by the scatter in the $f_l({\bf v})$ estimator. The fact that the dispersion increases as $\Ncell$ increases argues for using a small $\Ncell$. However, this conflicts with the need to ensure cells average over any substructures present in the halo. In general, we do not know, {\it a priori}, what substructures are present in a halo. However, we appeal to the fact that neither $\sigma_\R$ nor $s$ strongly depend on $\Ncell$ and that subhaloes typically have mass fractions of $\lesssim10^{-2}$ to determine an initial optimal $\Ncell$. By setting $\Ncell\sim 10^{-2}N_{\rm H}$, that is decomposing the halo into 128-1024 cells, we ensure that any compact substructures composed of $\lesssim10^{-2}N_{\rm H}$ will not significantly affect $f_{\rm bg}({\bf v})$. 

\par
\begin{figure}
    \centering
    \includegraphics[width=0.45\textwidth]{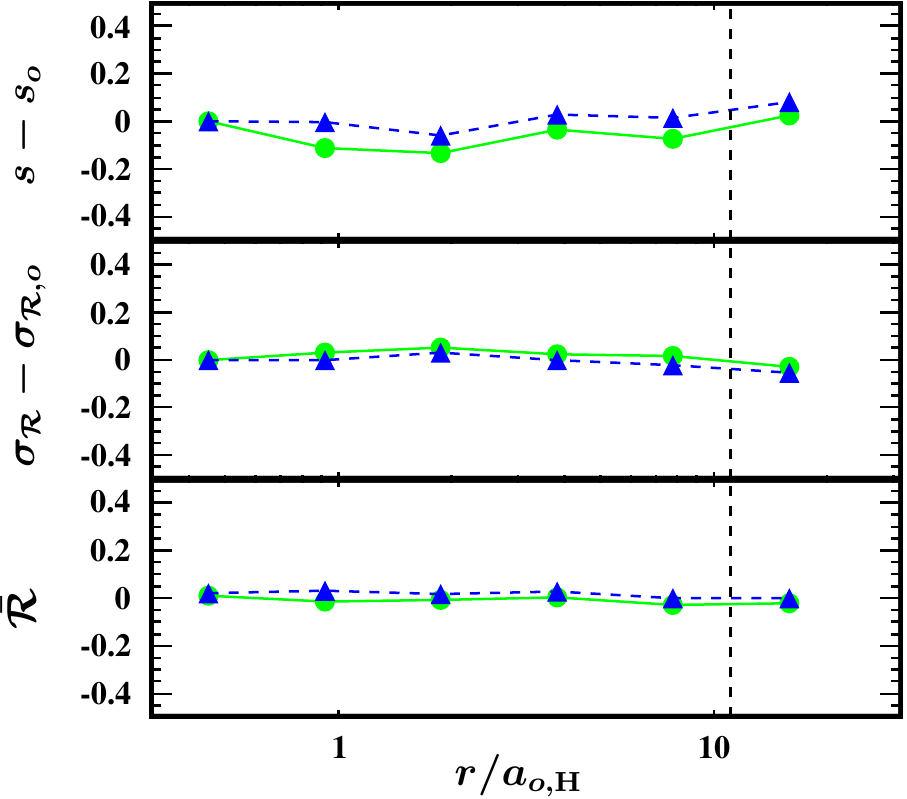}
    \caption{The variation in the parameters used to characterize the Skew-Gaussian $\R$-distribution as a function of radius for the radial bins shown in \Figref{fig:Rresdistrib} for $\Ncell=N_{{\rm cell},o}$ (filled circles, solid line) and $\Ncell=4N_{{\rm cell},o}=0.008N_{\rm H}$ (filled triangles, dashed line). As the mean should be centered on zero, we plot $\bar\R$ as a function of radius. For both $\sigma_\R$ and $s$, we plot the change relative to the quantity in the central radial bin, $\sigma_{\R,o}$ and $s_o$ respectively. We also indicate the virial radius with a dashed vertical line.}
    \label{fig:Rresdistrib-rad}
\end{figure}

As briefly mentioned before, we have found that in our test cases the $\R$ distribution does develop a systematic radial dependence for very large cell sizes, $\Ncell\gtrsim1\%N_{\rm H}$. This effect is shown in \Figref{fig:Rresdistrib-rad}, where we plot the parameters characterizing the local $\R$-distribution for the radial bins shown in the top panel of \Figref{fig:Rresdistrib} calculated using two different cell sizes, our fiducial value and a cell size four times as large ($\Ncell=0.8\%N_{\rm H}$). Both the mean and the variance do not significantly change even for radii near the virial radius nor do they display a strong radial trend. However, the skew does appear to develop a trend when using the larger cell size, increasing with increasing radius. It should be noted that though a systematic dependece is present, it is not strong even with this large a cell. This result indicates the effectiveness of our decomposition method.

\par
Our working hypothesis is that that changing the halo's bulk properties, such as is density profile, morphology or velocity anisotropy, will not change the form of the $\R$-distribution. We have tested our hypothesis using smooth haloes with different density profiles ranging from a cored isothermal profile to a steep $r^{-1.5}(1+r/a_{o})^{-1.5}$. These tests confirm the $\R$-distribution is insensitive to the density profile. We argue that changing other bulk properties such as the morphology will also leave the distribution unchanged. 

\par
Identifiable substructures will introduce secondary features in the $\R$-distribution that are anticipated to lie outside the scatter in $\R$ inherent in the smooth background. To account for this scatter, we estimate $\bar\R$ \& $\sigma_\R$. These quantities are calculated using the full width half max region about the most probable $\R$-value as this dominant peak should correspond to the smooth background. Once we have determined the mean and the dispersion, we calculate
\begin{align}
    \ELL_k=(\R_k-\bar{\R})/\sigma_{\R}.\label{eqn:ell}
\end{align}
So long as the $\R$-distribution is dominated by a singly peaked distribution, $\ELL$ is effectively a normalized deviation.

\subsubsection{Identifiable substructures}
To address what properties that will make a substructure appear dynamically distinct, let us consider a one dimensional Gaussian velocity that comprises of a substructure with mean $\bar v_{\rm s}$ and dispersion $\sigma^2_{\rm s}$ embedded in a background with $\sigma^2_{\rm bg}$. For a particle belonging to the velocity substructure, and ignoring the scatter in $\R$ for simplicity, 
\begin{align}
    \R=-\frac{(v-\bar{v}_{\rm s})^2}{2\sigma^2_{\rm s}}+\frac{v^2}{2\sigma^2_{\rm bg}}+\ln\left(\sigma_{\rm bg}/\sigma_{\rm s}\right),
\end{align}
where the first term corresponds to the local density, the second term to the estimated background density, and the third term is a ratio of the dispersions.

\par 
A particle belonging to this substructure should appear more strongly clustered in velocity space than the background and thus have $\R>0$, or equivalently $\ELL>0$. These particles will have velocities $v=\bar v_{\rm s}+a\sigma_{\rm s}$, where $a$ follows a normal distribution. For the case where $\bar v_{\rm s}=0$ a particle will appear dynamically distinct if
\begin{align}
    0&<-a^2\left(1-\frac{\sigma^{2}_{\rm s}}{\sigma^{2}_{\rm bg}}\right)+\ln\left(\sigma^2_{\rm bg}/\sigma^2_{\rm s}\right),\notag\\
    \sigma_{\rm s}^2/\sigma_{\rm bg}^2&<\exp\left[-a^2\left(1-\frac{\sigma^2_{\rm s}}{\sigma^2_{\rm bg}}\right)\right]
    \lesssim \frac{e^{-a^2}}{1-a^2e^{-a^2}}.
\end{align}
Therefore, to find a fraction $\erf(a/\sqrt{2})$ of the particles belonging to the substructure, the dispersion must be exponentially smaller. In the case where the dispersions are the same, a similar fraction will be identified so long as the offset between the mean velocities is significant, 
\begin{align}
    |\bar{v}_{\rm s}+a\sigma|>|a|\sigma.
\end{align}

\par
In the full three dimensions, the required difference in velocity or dispersions in any given dimension effectively decreases by factor of $\sim1/\sqrt{3}$. 

\subsection{Linking Outliers}\label{sec:linking}
\subsubsection{FOF-like algorithm}
Once the $\ELL$ distribution has been determined we apply a cut and only keep particles with $\ELL\geq\ELL_{\rm th}$. Since the $\ELL$ for a smooth halo follows an approximate Normal distribution, one could use $\ELLth\gtrsim2$ as this would eliminate $\gtrsim97.5\%$ of the background population in a smooth halo. By increasing $\ELLth$, we can decrease the likelihood of finding artificial groups but this comes at the expense of reducing the fraction of substructure that can be recovered. We will return to this issue in \Secref{sec:toymodels}. 

\par
The subset of particle determined through this cut is searched using a Friends-of-Friends-like algorithm where we link particles $i$ and $j$ iff
\begin{subequations}
\label{eqn:linkingcriteria}
%\begin{eqnarray}
\begin{gather}
    \frac{({\bf x}_i-{\bf x}_j)^2}{\ellx^2}<1,\\
    1/\Vr\leq  v_i/v_j\leq \Vr,\\
    \cos\Thetaop\leq \frac{{\bf v}_i\cdot{\bf v}_j}{v_i v_j},
\end{gather}
%\end{eqnarray}
\end{subequations}
where $\ellx$ is the physical linking length, $V_r$ is the velocity ratio, and $\cos\Thetaop$ is the velocity cosine. The first criterion is the standard \FOF\ criterion, linking particles that are separated by a distance less than $\ellx$. The last two criteria ensure that the particles have similar velocities. The reason the form of the velocity criteria differs from the configuration criterion, that is the reason we don't use $({\bf v}_i-{\bf v}_j)^2/\ellv^2<1$, is that tidal streams may have large velocities and dispersions. Consequently, scaling an allowed velocity dispersion, $\ellv^2$ is non-trivial. In total, this FOF algorithm has $4$ parameters, $\ELLth$, $\ellx$, $\Vr$ and $\cos\Thetaop$.

\par
As with all FOF algorithms, it is possible to find spurious groups with a poor choice of linking parameters and determining the optimal FOF parameters is not trivial. To guide the choices of these parameters we can consider either probabilistic or physical arguments. Consider first $\Vr$, which is simply a speed ratio. $\Vr\sim1$ would be a conservative choice whereas $\Vr\gg1$ will undoubtedly result in many spurious links. The related velocity parameter $\cos\Thetaop$ also has two limiting cases, the conservative value of $\cos\Thetaop\sim1$ and the relaxed condition of $\cos\Thetaop\sim-1$.

\par
The $\ellx$ linking-length parameter can significantly influence the results and, in the form used, there is no specific value to appeal to without prior knowledge of the halo structure. However, haloes of the virial overdensity $\Delta\rho_{\rm bg}$ are identified in cosmological simulations using $\ellx=\frac{2}{3}\Delta^{-1/3}$ times the inter-particle spacing. Equivalently, given a halo containing $N_{\Delta}$ particles in a radius $R_{\Delta}$, $\ell_{\rm x,H}=(2\pi/N_{\Delta})^{1/3}R_{\Delta}$. A natural choice to search for subhaloes would be $\ellx<\ell_{\rm x,H}$. However, we are also interested in finding physically diffuse tidal streams, therefore a reasonable choice is $\ellx\sim\ell_{\rm x,H}$.

\subsubsection{Removing artificial groups}
Regardless of the choice of parameters, it is always possible that random statistical fluctuations will give rise to artificial groups. In order to remove these artificial groups, we ensure that a group's $\ELL$ is significant relative to Poisson noise using a variation of a significance parameter as outlined in \cite{aubert2004}. Their {\sc adaptahop} algorithm identifies subhaloes by examining the physical density of particles and the groups thus found are required to have an average physical density that is significant relative to a threshold value when compared to Poisson noise. However, since our goal is to not only find physically overdense subhaloes but possibly physically underdense but dynamically distinct substructures, we cannot use density or phase-density. Instead we use the group's average $\langle\ELL\rangle$ value. If the group was purely artificial, one would expect $\langle\ELL\rangle$ to be within Poisson noise of the expected $\bar\ELL$ calculated using the background distribution and the threshold $\ELL_{\rm th}$ imposed. Thus, we require that a group composed of $N$ particles have an average $\langle\ELL\rangle$ that satisfies
\begin{align}
    \langle\ELL\rangle\geq\bar\ELL(\ELL_{\rm th})\left(1+\eta/\sqrt{N}\right).
\end{align}
Here $\eta$ is the required significance level and 
\begin{align}
    \bar\ELL=\frac{\int\limits_{\ELLth}^{\infty}xe^{-x^2/2}dx}{\int\limits_{\ELLth}^{\infty}e^{-x^2/2}dx}=\frac{\sqrt{\frac{2}{\pi}}e^{-\ELLth^2/2}}{1-{\rm erf}\left(\ELLth/\sqrt{2}\right)}.
\end{align}
We also require that a group must have $N\geq N_{\rm min}$. If the group does not satisfy the first criterion, the group member with the smallest $\ELL$ value is removed till the criterion is satisfied or until $N<N_{\rm min}$, in which case the group is removed. We generally set $\eta\gtrsim1$ and $N_{\rm min}=20$.

\section{Tests with a Single Subhalo}\label{sec:toymodels}
To test our \STF\ code and optimize the algorithm's parameters, we generate self-consistent equilibrium models of generalized NFW haloes with isotropic velocities using {\sc galactics}. The simulations are run using the parallel N-body tree-PM code \Gadget2 \citep{gadget2}.

\begin{figure*}
    \centering
    \includegraphics[height=0.66\textheight]{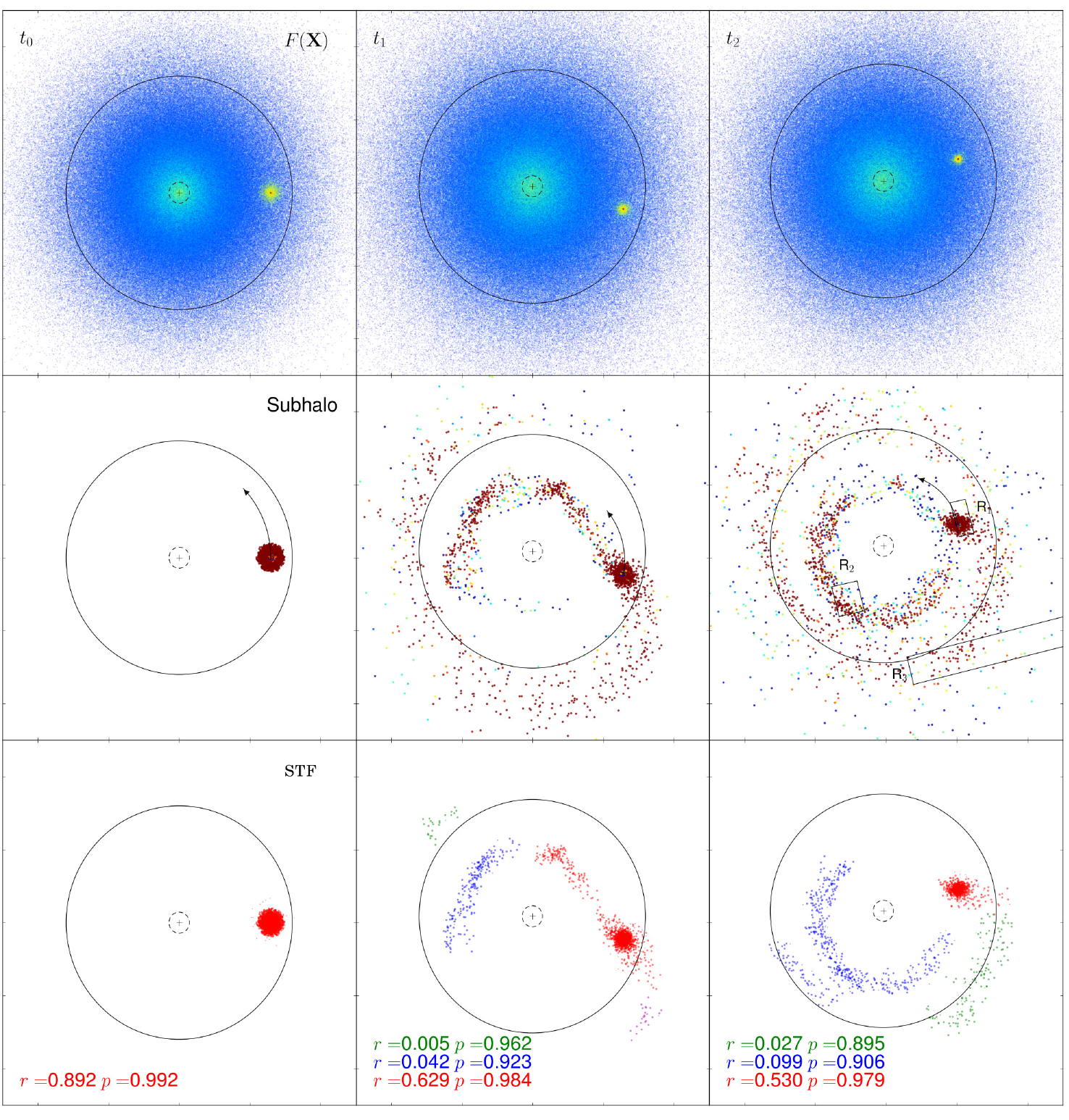}
    \caption{Projection of halo-subhalo system in the orbital plane for a box $2R_{\Delta}$ across at three different times, $t_o$ (left), $t_1$ (middle), and $t_2$ (right). The top row is a scatter plot of all the particles in the system where the particles of originating from the subhalo plotted over top the particles in the background. Particles coloured according to their estimated density in phase-space, ${\bf X}\equiv({\bf x},{\bf v})$, calculated with \citetalias{enbid} using a logarithmic colour scale spanning 7 dex, going from low to high density as one goes from dark blue to green to dark red. The middle row shows the physical distribution of subhalo particles colour coded according to their $\ELL$ value. In this row, dark blue corresponds to $\ELL\leq1$ and dark red corresponds to $\ELL\geq3$. The bottom row shows the group found using our algorithm with particles colour coded according to their group along with associated purity $p$ and recovery fraction $r$ of the three largest groups (see discussion below for information on the purity and recovery fraction). Also shown are the center-of-mass of the halo, a circle at $r=R_{\Delta}$ (solid), and another at $r=a_{o,{\rm H}}$ (dashed). The three rectangles denoted by R$_1$, R$_2$, and R$_3$ outline the three cells shown in \Figref{fig:singlesubellipELL}.}
    \label{fig:singlesubellip}
\end{figure*}
We first consider the case of a single subhalo that is stripped of particles by the tidal field of its (smooth) parent halo.  For this test, we assume a host halo identical to the one described in the previous section, that is $M_{\rm H}=10^{12}\Msun$, a characteristic velocity of $V_{o,{\rm H}}=344$~km/s and a scale radius of $a_{o,{\rm H}}=14.4$~kpc. We embed a subhalo with a mass equal to $0.005$ times the host halo mass. As with the host halo, the initial density profile of the subhalo is given by \Eqref{eqn:denprofile} with $\alpha=1$ and $\beta=3$. The structural parameters for the subhalo are $a_{o,{\rm S}}=2.6$~kpc, $V_{o,{\rm S}}=90$~km/s. The subhalo is placed on an circular orbit with radius of $r_{\rm apo, S}=9a_{o,{\rm H}}=0.8R_\Delta$ and an initial azimuthal period of $\approx4.6$~Gyr, ignoring the effects of dynamical friction. As a result of its orbital radius and structural parameters, the subhalo's tidal radius is $\approx3a_{o,{\rm S}}$. The subhalo is simply embedded in the larger host halo as \galactics\ cannot account for the presence of an external gravitational field when generating a halo. As a consequence of the diffuse nature of the subhalo, embedding it in the host causes particles in the very outer regions with retrograde velocities relative to subhalo's orbital velocity to become unbound.

\par
The evolution of this subhalo is shown in \Figref{fig:singlesubellip} at three different times: $t_0$, the start of the simulation; and $t_1$ and $t_2$ after two and four azimuthal orbits, respectively. After each azimuthal orbit, the subhalo loses about $\sim15\%$ of its mass in the form of diffuse, leading and trailing tidal streams. The subhalo is an easily identifiable phase-space peak in the top row which appears to decrease with time as the subhalo loses mass. In contrast, the stream is not visible on this scale, though it should be noted that much of the particles in the diffuse streams are denser than than their surroundings. This contrast between the tidal streams and the background is significantly enhanced once we look at their $\ELL$ value. We will return to this particular point later in \Secref{sec:ellvsphase}.

\subsection{Outlier Distribution}
\begin{figure*}
    \centering
    \includegraphics[width=0.85\textwidth]{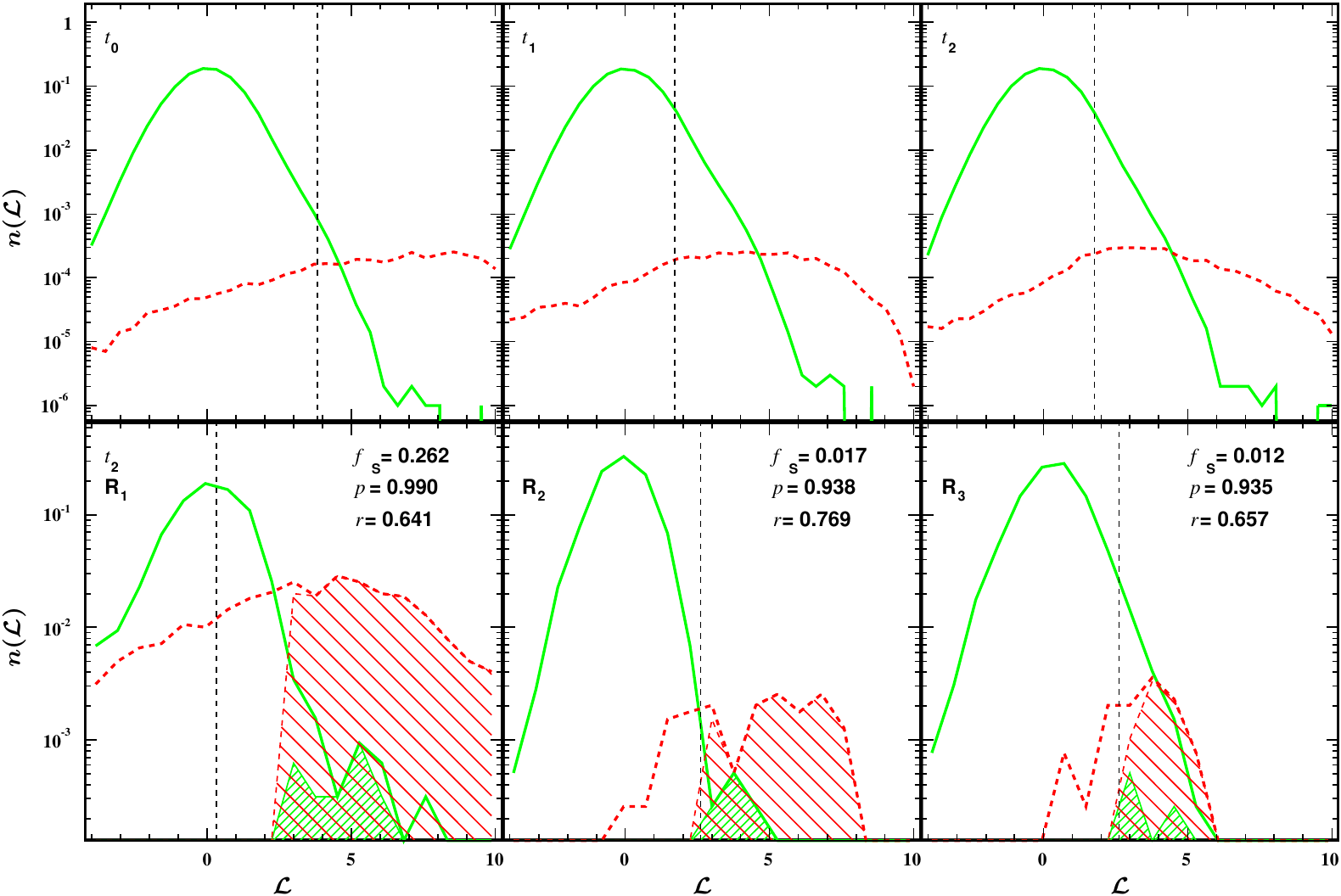}
    \caption{The normalized $\ELL$-distribution. The top panel shows the distribution of the entire halo-subhalo system at three different times calculated with our fiducial values for halo particles (solid line) and subhalo particles (dashed lines). The bottom panel shows the distribution in three specific cells at $t_2$ containing different fractions of subhalo particles $f_{\rm S}$. The filled regions in the bottom panel show the distribution of particles belonging to groups identified by our algorithm The region to the right of vertical dashed line contains $80\%$ of the subhalo population. We also show the purity and recovery fraction for each volume.}
    \label{fig:singlesubellipELL}
\end{figure*}
We use 256 cells with $\Ncell=3926$ (8-level tree) and $\Nv=32$ ($\Nse=32\Nv=1024$) to calculate the distribution of $\ELL$, the normalized deviations. In \Figref{fig:singlesubellipELL}, we plot the initial $\ELL$-distribution of the background and subhalo particles separately again at $t_0$, $t_1$ and $t_2$. This top row of the figure shows that more than $80\%$ of the subhalo population has $\ELL\gtrsim2$ at the times examined and is a small secondary feature in the global $\ELL$ distribution. Initially the two dynamical populations are well separated and as expected this distinction erodes over time as the subhalo is disrupted. At all times $\ELL\gtrsim4$ is dominated by the subhalo population. However, the separation between the two populations is greater than these results suggest since the subhalo particles are clustered in both configuration and velocity space. Halo particles with high-$\ELL$ values are generally statistical fluctuations and are thus unlikely to be clustered in either configuration or velocity space and therefore will not be linked.

\par
The middle row of \Figref{fig:singlesubellip} shows that the $\ELL$ values of the subhalo particles depend on where these particles reside in the tidal streams. Particles in the more diffuse outer edges of the stream tend to have smaller $\ELL$ values than those in the more densely populated portions of the tidal stream. This is not a surprising result as it is reflective of particles becoming more dynamically similar to the background. But perhaps more importantly, a particle's $\ELL$ value does not appear to depend strongly on the particle's radial position in the host halo. This is most clearly seen at $t_1$ where the densest portion of the leading stream, located at a $\sim5a_{o,{\rm H}}$, and the trailing stream, out past the virial radius, both have numerous particles with high $\ELL$ values. 

\par
The bottom row of \Figref{fig:singlesubellipELL} shows the $\ELL$-distribution at $t_2$ for three cells located at different points shown in \Figref{fig:singlesubellip}: R$_1$, located near in the progenitor subhalo and enclosing a substantial fraction of subhalo particles; R$_2$, containing subhalo particles primarily from a region along the leading tidal stream; and R$_3$ containing particles from the trailing tidal stream. In R$_1$ and R$_2$, the $\ELL$-distribution appears to be composed of two populations even though R$_2$ contains far fewer subhalo particles. In R$_3$ the two dynamical populations are not well separated. Without knowing {\it a priori} which particles originated from the subhalo, secondary features in the $\ELL$-distribution are not obvious. This poor separation may appear to indicate that $\ELL$ has a radial dependence since part of the reason for this poor separation is due to a larger number of background particles with high $\ELL$ values in R$_3$ compared to R$_1$ and R$_2$. However, most of the background particles with high $\ELL$ values originate from well outside the virial radius near the boundary of the system where the spatial sampling is very low. As a result of the poor spatial sampling, the local velocity density function estimator is no longer truly local and is strongly biased. In general however, different populations may have overlapping $\ELL$ distributions and an $\ELL$ cut-based approach that does not consider particles configurations is unlikely to find streams successfully. Hence our FOF algorithm is a necessary step that utilizes the fact that the subhalo particles are in fact clustered in both configuration and velocity space.

\subsection{Linking parameters \& groups identified}
We now turn to establishing the value of $\ELLth$. As noted earlier, based solely on the distribution observed for a smooth halo, a value of $\ELLth\gtrsim2$ would eliminate $\sim97.5\%$ of the background population from the search. The results shown in \Figref{fig:singlesubellipELL}, suggests this is a reasonable cut-off, but we use an even more conservative value of $\ELLth=2.8$, to exclude almost all the halo particles from the search. By looking for more significant outliers we can allow for larger velocity differences and physical distances. Our fiducial values are $\Vr=2$ and $\cos\Thetaop=0.97$. Due to the extended nature of the streams, we use a large linking length of $\ellx=10R_{\Delta}/N_{\rm H}^{1/3}\sim5\ell_{\rm x,H}=16~{\rm 
kpc}$, where the virial radius is defined using the overdensity of $\Delta=200$. 

\par
To assess the accuracy of our algorithm we calculate the purity and recovery fraction for the groups found \citep{enlink}. First we construct a classification matrix $A$ where each element $a_{i,j}$ represents the number of particles that belong to an intrinsic group $i$ that have been classified as belong to group $j$. We also note the cluster $j$ for which $a_{i,j}$ is a maximum and associate group $j$ with the intrinsic group $i$. We construct a recovery classification matrix $A^\prime$ such that $a^\prime_{i,j}=a_{i,j}$ if $j$ is the recovered group for intrinsic group $i$ and $a^\prime_{i,j}=0$ otherwise. The purity of an identified cluster $j$ is given by $p_j\equiv\sum_i a^\prime_{i,j}/ N_j$, where $N_j$ is the total number of particles associated with group $j$. The purity parameter $P$ is simply the mean purity. The recovery fraction is $r_i\equiv\sum_j a^\prime_{i,j}/ N_{i}$, where $N_i$ is the total number of particles associated with intrinsic group $i$. The recovery parameter is defined as the sum of every intrinsic groups recovery fraction, $R=\sum_i r_i$, and is anticipated to decay with time as particles phase-mix with the background halo. We also calculate the mean recovery $\bar{r}$. Together the quantities indicate how closely identified groups match intrinsic groups and the total number of recovered intrinsic groups found weighted by the fraction of that intrinsic group found. In this first example, there is only one intrinsic class.

\par
The groups found using these FOF parameters are shown in the bottom row of \Figref{fig:singlesubellip}. At all three times, the central subhalo is identified. Once the subhalo has formed tidal streams, a fraction of the tidal streams are also identified. At all times, the purity of each group is high, $\gtrsim0.9$, and the total recovery fraction for the intrinsic group is $\gtrsim0.6$. This fraction decreases with time due to two factors. Firstly, some portions of stream becoming sufficiently diffuse that the linking process does not identify them. Secondly, the natural phase-mixing of the subhalo particles with the background increases as time passes. Portions of the stream become very diffuse and though these particles are generally outliers, they are not linked with the $\ellx$ used. Furthermore, once a particle has been completely unbound from the subhalo, it will begin to phase-mix with the background as time passes. Arguably, as the phase-mixing occurs the recovery fraction should be normalized by a reduction in the size of the intrinsic subhalo class and our values can be seen an underestimates of the true recovery. We also note that some of the most distant outliers in the tidal tails are not linked simply due to the choice of linking parameters. There is no trivial solution to this linking length issue which we discuss later in \Secref{sec:iterativelinking}.

\par
At all times, only a very small number of halo particles are associated with the groups, $\sim100$ at $t_1$ and $t_2$. In order to asses whether these particles are truly false positives we compare the tangential and radial velocities of the ``misclassified'' halo particles at $t_1$ to the subhaloes particles in the same volume from $t_1$ to $t_2$. During this period, most of the halo particles lie well within the envelop of subhalo particles' orbital velocities. Comparing the distribution of velocities of the two populations using a Kolmogorov-Smirnov test. We find the time averaged probability that the halo particles belong to the same parent distribution of orbits as the subhalo particles to be $\sim0.2$ for both $v_r$ and $v_c$ over an entire orbital period between $t_1$ and $t_2$. Although the KS test is not very effective at measuring how similar two samples are and this probability is not high, given the low numbers, this comparison suggests that not all these particles are false positives and may have been swept into the subhalo. Approximately $\sim\%$ of these background particles lie within orbital space of correctly identified particles during this period. This in itself is an interesting result and worth further investigation, though a detailed analysis of how these background particles are swept up is beyond the scope of this work. 

\par
Returning to \Figref{fig:singlesubellipELL}, the recovery and purity fractions for the different cells provide useful information about the effectiveness of the linking process. Of most interest is the purity fraction in $R_2$ and $R_3$, which are located in the diffuse leading and trailing streams respectively.  The purity in these volumes is very high which is not what one would expect based solely on the observed $\ELL$-distribution, which might suggest a stronger inclusion of non-subhalo particles. Again, most of the subhalo particles above $\ELLth$ are linked whereas a substantial fraction of the halo particles above $\ELLth$ are not. These halo particles remain unlinked due to the fact their large $\ELL$ values are either due to statistical fluctuations or biased estimates for particles lying well outside the virial radius at large physical distances from other particles, and, unlike the subhalo particles, are not clustered in phase-space.

\subsection{Optimizing the choice of $\ELLth$}
\begin{figure}
    \centering
    \includegraphics[width=0.45\textwidth]{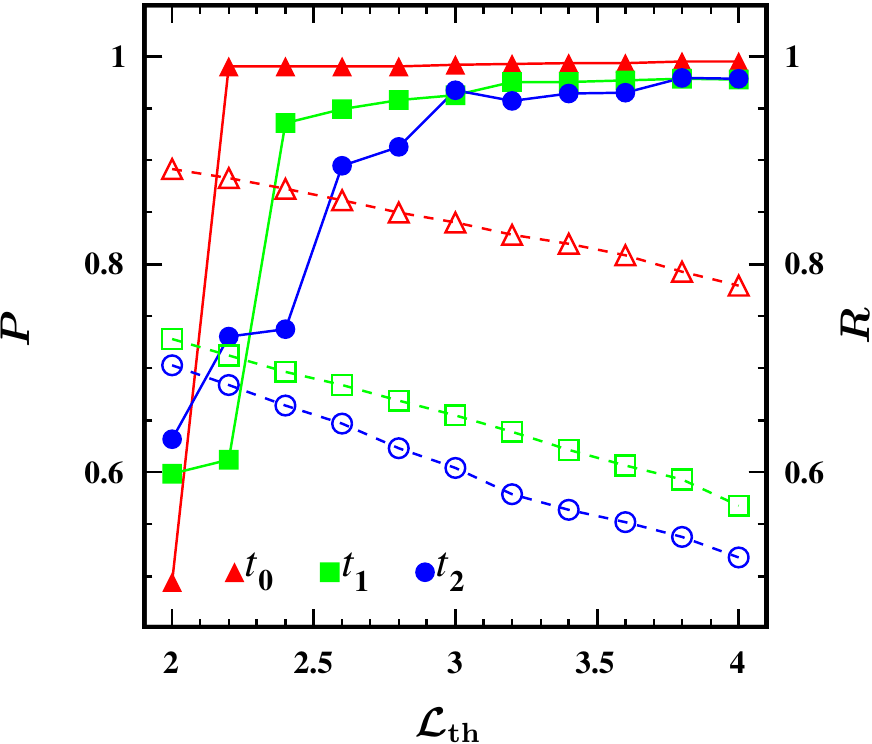}
    \caption{The mean purity (filled points) and total recovery fraction (open points) as a function of $\ELLth$ at three different times.}
    \label{fig:purityrecoveryellip}
\end{figure}

Figure \ref{fig:singlesubellipELL} also shows that there are subhalo particles below the threshold used. No choice of $\ELLth$ is perfect and we must tolerate the inclusion of some background along with the loss of substructure particles. Our fiducial choice of $\ELLth$ was based on probabilistic arguments and investigation of the $\ELL$ distribution. A more rigorous determination of the optimal $\ELLth$ requires searching for the value that maximizes both $P$ and $R$. We show the dependence of these quantities on the threshold in \Figref{fig:purityrecoveryellip} at three different times. Initially, one can use a low threshold and still recover a large fraction of the subhalo with a high purity. Once the tidal streams have formed, the purity is relatively insensitive to the threshold used for $\ELLth\gtrsim2.5$ and converges to $\approx0.95$ for $\ELLth\geq3$. The differences in $P$ between $t_1$ and $t_2$ for $\ELLth\gtrsim2.5$ are small, especially considering that as the system evolves, the subhalo will have swept up more of the background particles thereby resulting in $P$ at $t_2$ being systematically underestimated relative to $P$ at $t_1$. The recovery fraction decreases as we increase $\ELLth$ at all times and the sensitivity of $R$ on $\ELLth$ increases slightly with time as more of the subhalo's particles mix with the background. The overall decrease in $R$ at a given $\ELLth$ with time is due to the increasing amount of particles that are truly lost from substructure.

\par
Based on this model, $\ELLth\sim2$ is clearly enough to ensure a large fraction of a subhalo is found with a very high purity at $t_0$, but that is not the case once tidal streams have formed. Since a true cosmological halo at any given instant will have both subhalos and streams, it is more constructive to look at $P$ and $R$ at $t\geq t_1$. At these late times there are diffuse streams that are physically well separated from their progenitor subhalo. The choice of $\ELLth$ that maximizes $P+R$ at these late times is $\sim2.8$, our fiducial choice. This optimal threshold appears to hold true regardless of a stream's orbit, whether circular or radial.

\subsection{Merging candidate groups to increase the recovery fraction} \label{sec:iterativelinking}
While recovery fractions are generally high, streams are never perfectly recovered. The unlinked particles lie in the most diffuse regions of the tidal stream, and although many of these particles are outliers, they are separated by distances larger than the linking-length $\ellx$. The presence of diffuse regions in the stream has also lead to the algorithm splitting the tidal stream into several groups at $t_1$ and $t_2$. While this splitting might seem to be cause for concern, we again emphasize it is primarily due to the very diffuse nature of certain portions of the tidal stream. For example the largest group found at $t_2$, which is composed of $151$ particles, has $r=0.027$, spans approximately $\sim200$~kpc and accounts for $\sim7\%$ of the mass in the volume occupied by this group. The closest particles belonging to the neighbouring group, which contains the subhalo, lie just outside the velocity and spatial criteria used.

\par
The recovery fraction can be increased by lowering thresholds and increasing linking lengths, but in turn this will reduce the purity. An adequate balance between the two must be chosen. Since tidal streams can be very diffuse and pass through one another there is no simple adaptive way of choosing a spatial linking length. A more robust iterative approach would use conservative values for the FOF parameters, find an initial set of candidate groups, relax the FOF criteria, generate a new subset and then search this new set for particles that meet our FOF criteria with previously identified candidate particles.

\par
We have found that this method works quite well, effectively merging several of the groups together, increasing the recovery fraction of the particles in the tidal streams by a factor of 2 while keep the purity $\gtrsim0.90$. The net recovery increases from $0.68$ to $0.75$ and $0.65$ to $0.75$ at $t_1$ and $t_2$, respectively, keeping in mind that a majority of the particles still lie in the subhalo. However, we caution that this model is a best case scenario for this approach as there is only one substructure. Realistic haloes will contain numerous substructures which may overlap in configuration and velocity space and using the iterative method may merge substructures together. 

\subsection{The $\ELL$ and phase-space density contrast of tidal streams}\label{sec:ellvsphase}
\begin{figure*}
    \centering
    \includegraphics[width=0.325\textwidth]{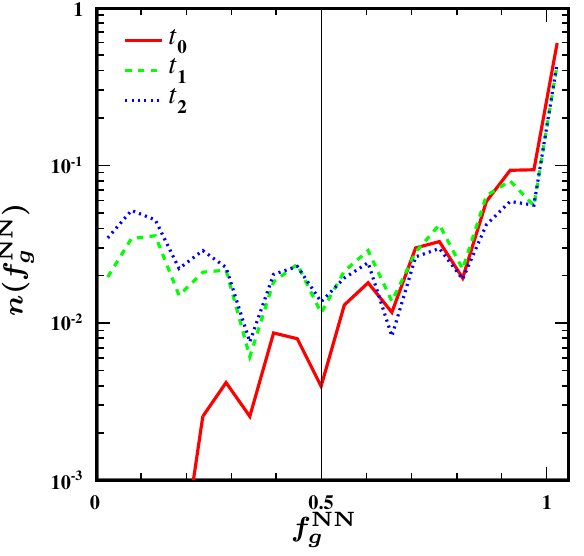}
    \includegraphics[width=0.325\textwidth]{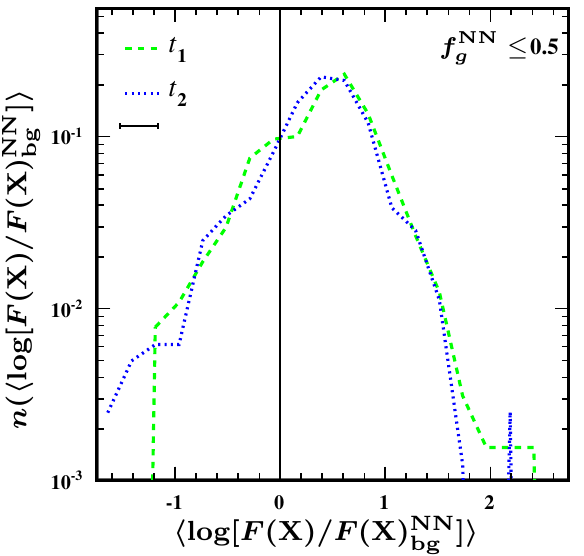}
    \includegraphics[width=0.325\textwidth]{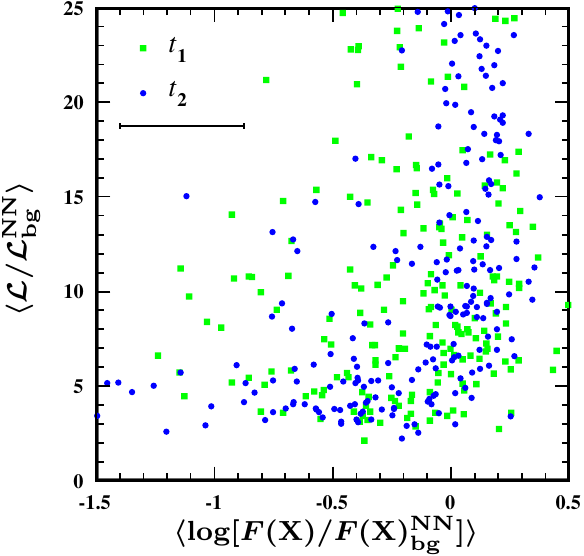}
    \caption{Comparison of correctly identified group members with nearest neighbours in phase-space at three different times as seen in \Figref{fig:singlesubellip}. Left: The normalized distribution of $f^{\rm NN}_g$ for all correctly identified group members along with a vertical line at $f^{\rm NN}_g=0.5$. Middle: The normalized distribution of the average phase-space density contrast with respect to the background, $\logphase$, of correctly identified group members with $f^{\rm NN}_g\leq0.5$. Also shown is a solid vertical line at $\logphase=0$ and a typical error bar for $\logphase$. In these panels, all the distributions are normalized individually. Right: $\logphase$ versus the average contrast in the normalized standard deviation with respect to the background, $\langle\ELL/\ELL^{\rm NN}_{\rm bg}\rangle$, of particles with $f^{\rm NN}_g\leq0.5$ and $\logphase-{\rm SD}\logphase\leq0$. Again we show a typical error bar for $\logphase$.}
    \label{fig:phasenncompellip}
\end{figure*}
Numerous studies show that the contrast between the outer region of a subhalo relative to the background is much greater in phase-space than in physical space \citep{enbid,elahi2009,helmi2000,enlink,hsf}. Algorithms such as \enlink\ and \hsf\ use this increased contrast to identify these substructures. However, accurately computing the local phase-space density is a non-trivial task. In the following discussion, we show that the difficulty of accurately computing the phase-space density means that these algorithms might not find the extended tidal streams even though these structures are dynamically distinct.

\par
In figure \ref{fig:phasenncompellip}, we compare the phase-space structure of the subhalo particles relative to the background. To compare the properties of correctly identified group members with the background, we first find the nearest $N_{\rm NN}$ neighbours in phase-space for each group member. Here we set $N_{\rm NN}=32$. The key to correctly identifying near neighbours in phase-space lies in determining the optimal locally adaptive metric describing the local phase-space volume. We follow the method outlined in \citetalias{enbid} to find these nearest neighbours.

\par
Next, for each correctly identified group member we determine $f^{\rm NN}_g$, the fraction of nearest phase-space neighbours of a group member that are also members of the group as shown the left panel of \Figref{fig:phasenncompellip}. Prior to the formation of tidal tails, most group members' nearest phase-space neighbours are also members of the group. The initially small fraction of particles with $f^{\rm NN}_g\sim0.5$, $\approx0.03$, reside in the outer edges of the subhalo. As the subhalo begins to lose mass and form tidal tails, this fraction grows to $\approx0.20$ and $\approx0.25$ at $t_1$ and $t_2$ respectively. The anisotropic velocity structure and diffuse configuration of tidal streams makes accurately determine nearby neighbours nontrivial and consequently linking nearest neighbours is not the ideal method for identifying these structures with a high purity.

\par
We compare the estimated phase-space density of particles located in the tidal tails at $t>t_o$, which have $f^{\rm NN}_g\leq0.5$, to their nearest phase-space neighbours in the background population by calculating the average logarithmic phase-space density contrast, $\logphase$. The distribution of $\logphase$ at $t_1$ and $t_2$ is shown in the middle panel of \Figref{fig:phasenncompellip}. Although most of the particles in the tidal streams are denser than their surroundings, $\approx20\%$ are less dense. If we account for the local scatter by calculating the standard deviation ${\rm SD}\logphase$, the fraction of particles that are not significantly denser than their surroundings, that is particles with $\logphase-{\rm SD}\logphase\leq0$, increases to $\approx25\%$.

\par
For this low density subset, we determine $\langle|\ELL/\ELL^{\rm NN}_{\rm bg}|\rangle$, the average ratio of the particle's $\ELL$ value relative the nearest phase-space neighbours belonging to the background. The right panel of \Figref{fig:phasenncompellip} shows that these particles typically have much larger values despite appearing underdense. It is perhaps significant that these quantities do not appear strongly correlated for the particles residing in the diffuse tidal streams. 

\par
These results indicate that particles in tidal streams may have negligible contrast in their estimated phase-space density relative to their surroundings due to their diffuse configuration and the inherent noise in the phase-space density estimator. Consequently, any search that used only phase-space density may miss identifying particles residing in the very diffuse portions of the stream while ensuring a high purity whereas our algorithm will be able to identify them.

\subsection{Other single subhalo tests} 
We have examined the impact of placing the subhalo on different orbits. In all cases, regardless of the subhalo's orbit, substructure was recovered with purity and recovery fractions of $\gtrsim0.8$ and $\gtrsim0.6$. For example, we analyzed a subhalo that was on a primarily radial orbit plunging through the center of the host with a peri- and apocenter of $r_{\rm peri, S}=0.1a_{\rm H}$ and $r_{\rm apo, S}=7.5a_{\rm H}=0.67R_\Delta$, respectively. Over the period analyzed, corresponded to several radial orbits during which the subhalo lost $\sim60\%$ of its bound mass, the subhalo's $\ELL$-distribution is similar to that of the subhalo on an elliptical orbit. That is the substructure distribution is a small secondary feature in the global $\ELL$ distribution and dominates the distribution for $\ELL\gtrsim4$, though it is more strongly peaked than the distribution in th elliptical orbit. Over the period examined, our algorithm identified a single group containing both the progenitor subhalo and the tidal streams and shells produced by the subhalo using the same fiducial threshold. The purity of the group was always $\gtrsim0.95$ while the recovery fraction dropped to $r\approx0.8$ after starting at $r\gtrsim0.9$. 

\par
It is interesting to note that in this example the false positives identified have orbits that do not strongly differ from the orbits of the particle in the substructure. Approximately $70\%$ of these false positives remain in the orbital space of the subhalo over one radial period. Again, it appears that some of the background has been swept up by the subhalo. Though it is beyond the scope of this work, it would be interesting to examine in detail this ``mass accretion'' and its dependence on the subhalo's orbit and position.

\par
These tests taken together show that a particle's $\ELL$ value does not strongly depend on the particles position in the host halo itself or its orbit, but rather on whether the particle belongs to a velocity substructure. Hence, our algorithm is ideally suited to finding substructure since $\ELL$ is an excellent discriminator. 

\section{Towards a realistic model}\label{sec:halomany}
\begin{figure*}
    \centering
    \includegraphics[width=0.90\textwidth]{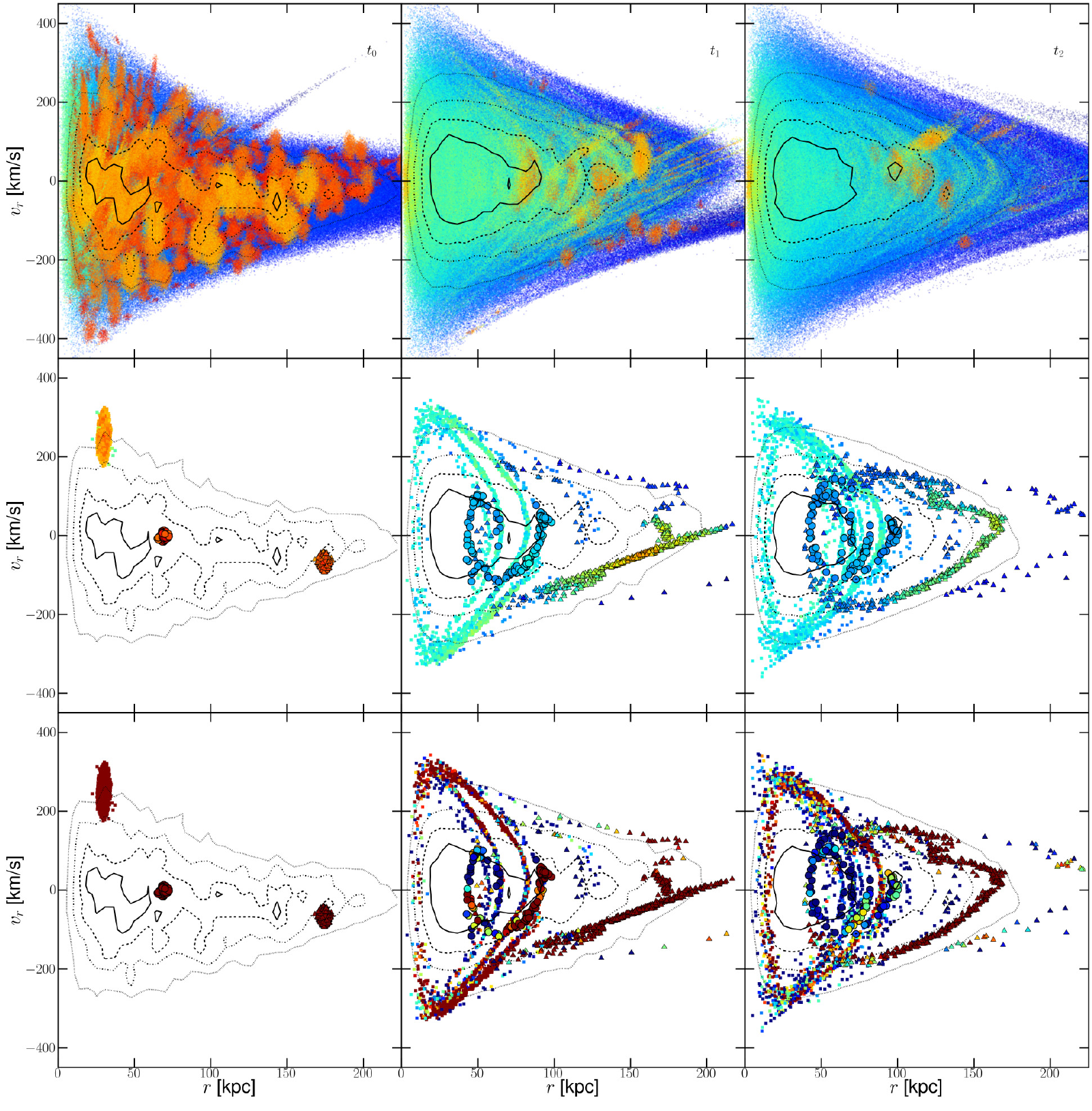}
    \caption{Phase-space $r-v_r$ plot of halo at three different times, $t_o$, $t_1$, $t_2$ (top) and the evolution of three subhaloes (bottom two rows). Particles in the first two rows are colour coded according to their phase-space density estimated with \citetalias{enbid} using the same logarithmic colour scale as the top row of \Figref{fig:singlesubellip}. In the top row particles belonging to substructure are plotted over top of particles in the background. Particles in the bottom row are colour coded according to their $\ELL$ value using the same colour scale as the middle row of \Figref{fig:singlesubellip}. Also shown are mass contours.}
    \label{fig:halomany-phase}
\end{figure*}
\subsection{Constructing the test case and general evolution}
We next consider a halo with a large number of subhaloes. We have constructed a halo, as outlined below, where we know, {\it a priori}, where all the particles come from. The tests here serve as an intermediate step towards full cosmological haloes. 

\par
To construct the test system, subhaloes were given cored isothermal density profiles with mass fractions from $f=[5\times10^{-5}$,$5\times10^{-3}]$ and were drawn from a mass distribution such that the number of subhaloes per logarithmic mass fraction is $dn/d\ln f=Af^{-0.9}$, where power-law index is the one commonly observed in cosmological simulations (e.g.~\citealp{moore1999,diemand2008,springel2008,elahi2009}). These studies also show that the typical amount of the host's mass in bound subhaloes in mass fraction range of $f\sim10^{-6}-10^{-2}$ is $\sim10\%$. To follow a large number of haloes at our chosen mass resolution we set the mass fraction to a much higher value of $30\%$ of the host's mass yielding $1496$ subhaloes in total. We placed subhaloes by replacing particles of the host halo with radii from $r=[a_{o,{\rm H}},R_{\Delta}]$ with subhaloes and set the subhalo's center-of-mass and center-of-mass velocity to the replaced particle's position and velocity. The resulting system has a total mass of $1.6\times10^{12}~\Msun$ and is composed of $1.42\times10^6$ particles. This system is evolved for $10$~Gyr, corresponding to $\sim$5 dynamical times for the halo at the virial radius.

\par
The phase-space structure of the halo is shown in \Figref{fig:halomany-phase} at three different times, a few Myr after the start of the simulation, $5$~Gyr later, and after $10$~Gyr. We also highlight the evolution of three subhaloes in the bottom two panels of \Figref{fig:halomany-phase}. Initially, there are numerous regions in halo's phase-space that contain particles with high phase-space density, corresponding to dense dynamically cold subhaloes. Particles with these high phase-space densities account for $30\%$ of the halo's mass initially and typically have $\logphase\approx3.5$. After $5$~Gyr, only $\approx4\%$ of the particles have similarly high phase-space densities.

\subsection{Evolution of subhaloes}
The smallest subhalo shown, (circles), is composed of $\approx100$ particles and is initially on an approximately circular orbit with an orbital period of $3.7$~Gyr. After $5$~Gyr the particles of this subhalo occupy a large volume in phase-space and no longer have a comparatively high phase-space density relative to the background. The mean logarithm physical and phase overdensity are $\logrho=0.01$ and $\logphase=0.35$ respectively, where the physical overdensity is computed relative to nearby physical neighbours. After $10$~Gyr, the mean $\logrho\approx0$. Despite this low contrast and the fact that only $\approx35\%$ of the subhalo particles are denser than their surroundings by a factor of one standard deviation, $\approx40\%$ of the subhalo's particles have $\ELL\gtrsim3$. This comparison indicates that we are capable of picking out $14\%$ more of this substructure based on $\ELL$ compared to $f({\bf X})$ at a greater normalized contrasted, which should result in a higher purity.

\par
The second subhalo shown, (triangles), is composed of 600 particles and is on a primarily radial orbit. The subhalo's core is still relatively intact till after $10$~Gyr and it has generated extended tidal streams over this period. At $t_1$, the fraction of the particles in this group with $\logphase-{\rm SD}\logphase>0$ is $\approx0.7$ and the mean $\logphase$ for the group is $\approx1.4$. By $t_2$ the fraction of overdense particles is $\approx0.5$ and the mean $\logphase$ of the subhalo has dropped to $\approx0.4$. At all times, the particles in the tidal streams also have high $\ELL$ values, even in cases where the particle has a low phase-space density.

\par
The most massive subhalo shown, (squares), is composed of $\approx2000$ particles and is on an elliptical orbit. Although this subhalo appears to be completely disrupted after $5$~Gyr or $\approx2.5$ orbits and has a mean $\logphase\approx0$, $\approx50\%$ of the subhalo's particles reside in a specific region in the $r-v_r$ plane and still have comparatively higher phase-space densities than the background. After $10$~Gyr, the fraction of particles with $\logphase-{\rm SD}\logphase>0$ is $\approx0.26$ though the fraction of these particles with $\ELL\gtrsim3$ is $\approx0.30$. 

\par
Figure \ref{fig:halomany-phase} highlights the fact that disrupted subhaloes can leave distinct high phase-space density features in the $r-v_r$ plane, though the phase-space density contrast of these features decreases with time. More importantly, this figure shows that even when the estimated contrast in phase-space may be negligible after several orbits, the particles originating from dynamically distinct substructures still appear to be velocity outliers.

\subsection{Substructures Identified}
\begin{figure*}
    \centering
    \includegraphics[width=0.3275\textwidth]{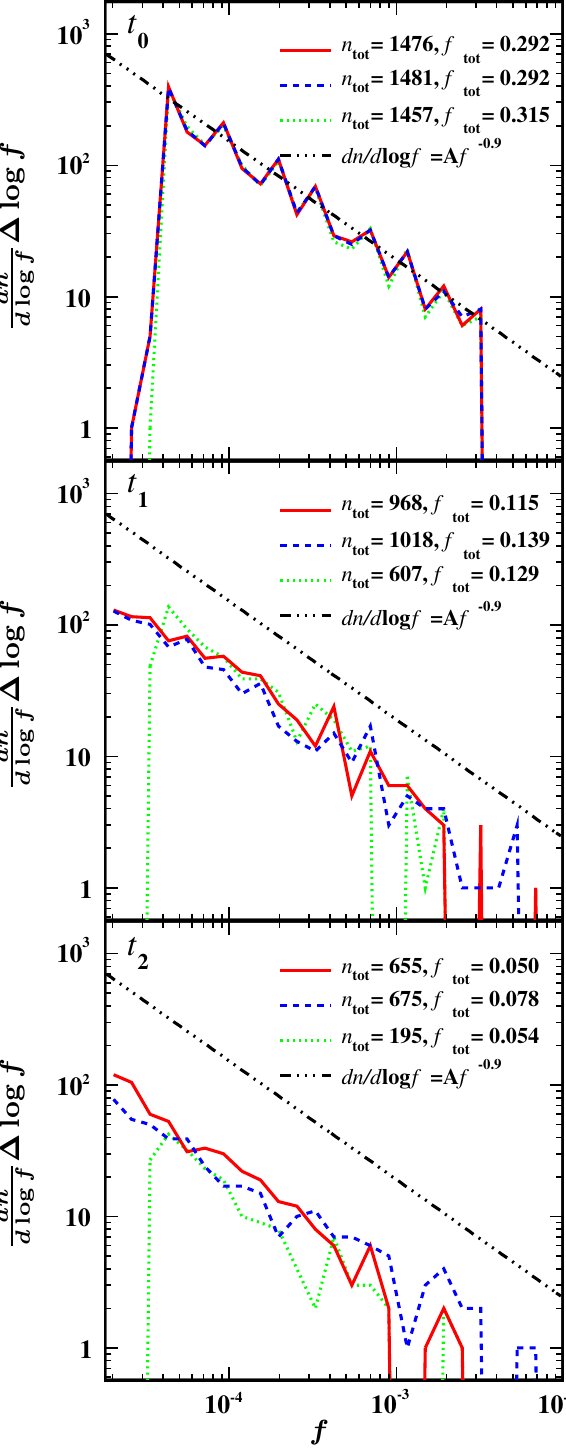}
    \includegraphics[width=0.3275\textwidth]{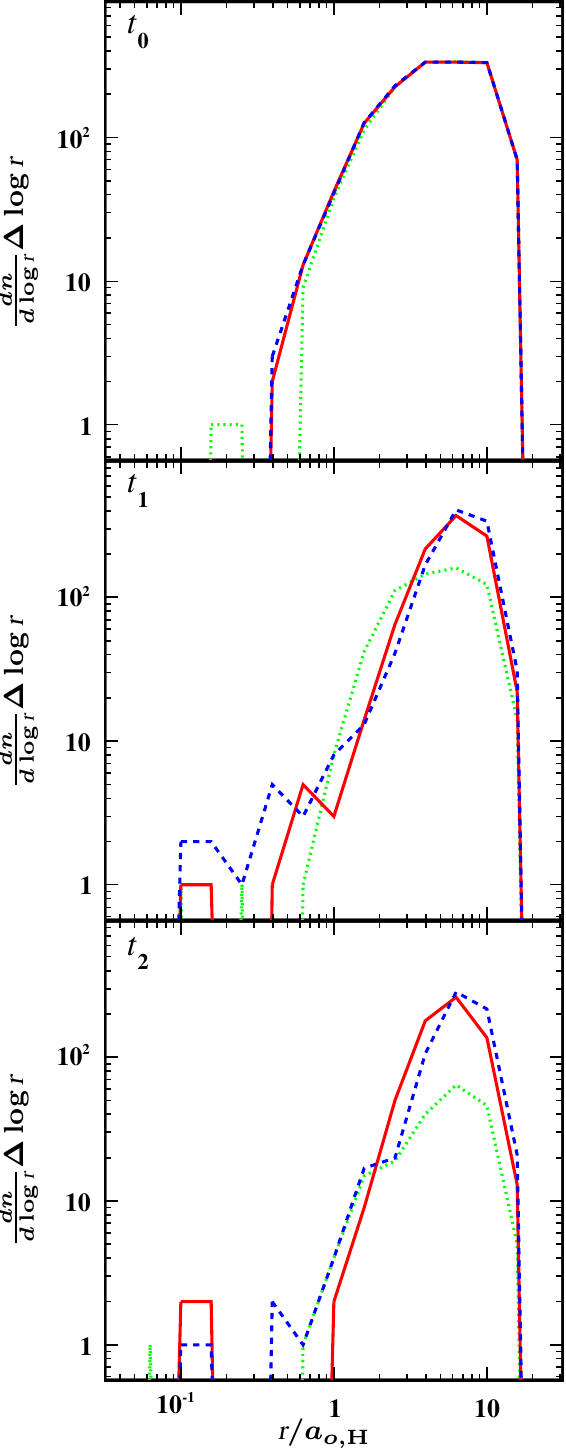}
    \includegraphics[width=0.3275\textwidth]{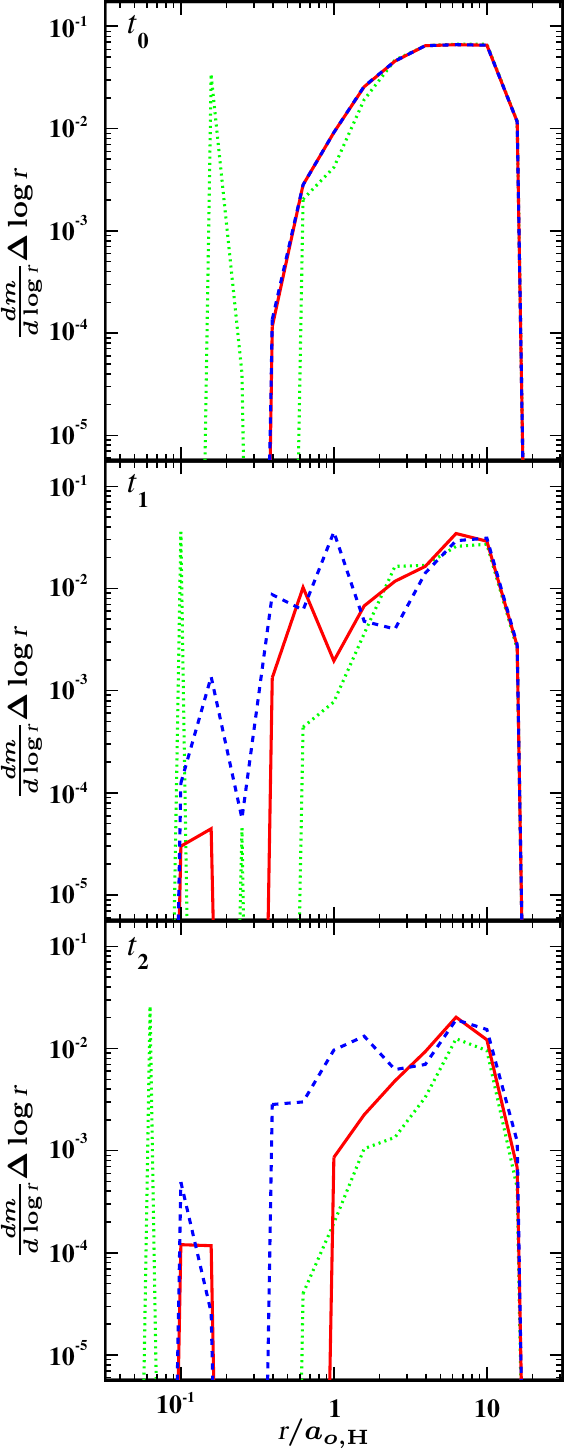}
    \caption{Distributions of groups identified at three different times for three different searches, \STF-1.5 (solid), \STF-IT (dashed), and \6DFOF\ (dotted):  substructure mass function $dn/d\log f$ (left);  radial distribution $dn/d\log r$ (middle); and radial mass distribution $dm/d\log r$ (right). The radial position of a group is given by ${\text{\slshape r}}=|x_{{\rm cm},i}-X_{\rm cm,H}|$. Also shown in the left panel is a dashed-dotted line indicating the mass distribution used to generate the initial conditions. We also show the total number of groups found and total mass fraction found in substructures using each algorithm.}
    \label{fig:halomany-massfunc}
\end{figure*}
To calculate the normalized deviations we again use $\Nv=32$ and 256 cells with $\Ncell=5530$. We use $\ELLth=2.8$, $\Vr=2$, $\cos\Thetaop=0.97$, and $\ellx\approx1.5\ell_{\rm x,H}=5$~kpc as our fiducial values. We also test our iterative method where initial candidate groups are found using the fiducial values and $N_{\rm min}=10$, then the system is searched again with $\ELL$ reduced by $10\%$, $\Vr$ and $\Thetaop$ increased by $10\%$, and $\ellx$ is increased by $50\%$ and $N_{\rm min}=20$. We will denote these two searches as \STF-1.5 and \STF-IT. For comparison, we use a \6DFOF\ algorithm to find regions of high phase-space density with $\ellx=\ell_{\rm x,H}$ and $\ell_{\rm v}=0.25 V_{c,\Delta}$, where $V_{c,\Delta}^2=GM(R_{\Delta})/R_{\Delta}$ is the circular velocity at the virial radius. The \6DFOF\ parameters are chosen by examining the phase-space contrast of several randomly chosen subhaloes and ensuring that the \6DFOF\ search identifies $\gtrsim95\%$ of the subhaloes at $t_o$. We also set the minimum number of particles for the \6DFOF\ search to 50, otherwise the \6DFOF\ algorithm at $t_0$ finds spurious groups.

\par
In \Figref{fig:halomany-massfunc}, we show mass and radial distributions of the groups identified by each search. The mass distribution power law at $t_0$ is reproduced by all three algorithms, but as the system evolves tidal disruption reduces the number of substructures identified. Accordingly, the total fraction of mass in substructure, $f_{\rm tot}$, also decreases with time.  Initially, \6DFOF\ finds roughly the same number of groups as \STF, but as subhaloes are tidally disrupted it finds progressively fewer groups compared to \STF. At $t_2$, after subtracting out the contribution from the halo core that is falsely identified by \6DFOF\, the mass fraction in substructure it identifies is $20-50\%$ less than that found with the \STF\ searches. It also finds signficantly fewer structures than the \STF\ searches.

\par
The radial distribution indicates that the number of velocity substructures within a few scale radii steadily decreases with time. This is not unexpected since the subhaloes that cross the very central region of the host experience the strongest tidal and impulsive forces and are more likely to be disrupted. The apparent dearth of substructure is in qualitative agreement with other studies \citep{springel2008,diemand2008}. However, the radial distribution for the \STF\ groups suggests that the central regions is not completely devoid of substructure.

\par
The overall performance of each search is summarized in \Tableref{tab:halomany-summary}. The random placement of the subhaloes means that two subhaloes may overlap in both configuration and phase-space. Consequently, we define the intrinsic class of a particle using the initial results from each group finder to determine the recovery fraction. All find $\gtrsim1450$ substructures. The small initial differences between the searches are due to a few low mass subhaloes lying in close proximity to another subhalo in phase-space and being linked to it.
\begin{table}
\centering
\caption{Performance of \STF\ and \6DFOF}
\label{tab:halomany-summary}
\begin{tabular*}{0.45\textwidth}{@{}l@{}c@{}rccc}
%\begin{tabular}{lcccc}
    \hline\hline
    & Time & $N_{\rm g}$ & ${P\pm\sigma_{P}}\;_{\rm min}^{\rm max}$ & $R$, $\bar{r_i}\pm\sigma_r\;_{\rm min}^{\rm max}$\vspace{2pt}\\\hline
    \multirow{3}{*}{\vspace{-5pt}\STF-1.5}& $t_o$ & $1476$ & & \vspace{2pt}\\
    & $t_1$ & 968 & $0.96\pm0.10\;_{0.19}^{1.00}$ & $267.5$, $0.18\pm0.27\;_0^{1.00}$ \vspace{2pt}\\
    & $t_2$ & 655 & $0.93\pm0.12\;_{0.31}^{1.00}$ & $77.2$, $0.05\pm0.14\;_{0}^{0.99}$ \vspace{2pt}\\\hline
    \multirow{3}{*}{\vspace{-5pt}\STF-IT}& $t_o$ & $1481$ & & \vspace{2pt}\\
    & $t_1$ & 1018 & $0.96\pm0.10\;_{0.11}^{1.00}$ & $306.6$, $0.21\pm0.28\;_0^{1.00}$ \vspace{2pt}\\
    & $t_2$ & 675 & $0.93\pm0.14\;_{0.27}^{1.00}$ & $108.9$, $0.07\pm0.17\;_{0}^{0.99}$ \vspace{2pt}\\\hline
    \multirow{3}{*}{\vspace{-5pt}\6DFOF} & $t_o$ & $1457$ & & \vspace{2pt}\\
    & $t_1$ & 607 & $0.92\pm0.15\;_{0.20}^{1.00}$ & $171.4$, $0.12\pm0.23\;_{0}^{1.00}$ \vspace{2pt}\\
    & $t_1$ & 195 & $0.86\pm0.21\;_{0.20}^{1.00}$ & $38.4$, $0.03\pm0.12\;_{0}^{0.99}$ \vspace{2pt}\\
\end{tabular*}
%\end{tabular}
\end{table}

\par
The mean purity of the groups identified by \STF\ is $\gtrsim0.90$, which is slightly higher than \6DFOF. The dispersion in the purity is also smaller for \STF\ and shows little evolution in time while \6DFOF\ shows a significant increase in dispersion at the final time. This increase can be attributed to the increasing number of streams at later times and the inability of \6DFOF\ to correctly identify them. The minimum purity found in  the \STF\ searches is a product of streams overlapping rather than incorrect identification of background particles. Since the purity of a group is calculated by determining the dominant intrinsic class of a group, if several subhaloes are on similar orbits and merge and produce tidal tails, the calculated purity will low. For example, the group with the minimum purity identified by \STF-IT at $t_2$ is a massive substructure resulting from the merger of several subhaloes and several overlapping tidal streams. Only $3\%$ of the particles belonging to this group were initially part of the halo background at $t_0$.

\par
\STF\ recovers $0.45$ of the smallest subhalo highlighted in \Figref{fig:halomany-phase} at $t_1$ even though it appears to be disrupted, although by $t_2$ none of the subhalo's particles are recovered. A similar fraction of the massive subhalo in this figure is recovered at $t_1$. After 10~Gyr only $\sim0.17$ is recovered, despite the fact that the subhalo has been tidally disrupted and occupies a large physical volume while only accounting for $\approx5\%$ of the mass in this volume. The highlighted subhalo with the extended radial tidal tails is identified at $t_1$ and $t_2$ with recovery fractions of $0.8$ and $0.6$, respectively.

\par
The difference between \STF-1.5 and \STF-IT suggests that the iterative method is a significant improvement over the standard method. The recovery fraction increases by $14\%$ and $40\%$ at $t_1$ and $t_2$, respectively, while $P$ is not significantly decreased. However, there is a significant decrease in the minimum purity of \STF-IT at $t_1$. This decrease is due to one instance where several groups have been merged together, including the most massive group found by \STF-1.5. The increase in $R$ and decrease in $P$ using the iterative method depends on how significantly the FOF parameters are changed after the initial search. To determine an optimal fractional change in the FOF parameters, we have used the \STF-1.5 search as a baseline and varied the change which increases $R$ while keeping $P$ unchanged. We find larger changes than those used here degrade the mean purity and increase the instances of high mass, low purity groups that are the result of several substructures being linked together.

\par
In summary, the recovery fraction of the \STF\ search is greater than that of the \6DFOF\ search by $\gtrsim2$. Furthermore, the mean recovered fraction for each group is also significantly higher in the \STF\ searches. Increasing $\ellx$ used by \STF\ increases the recovery fraction, particularly of low mass tidally disrupted groups located at large radii without significantly reducing the purity. For all three algorithms the recovery fraction is $\lesssim1/4$ times the initial number of groups found, due to subhaloes being disrupted and phase-mixing with the background.

\subsection{Subhalo or Stream?}
Substructures can range from relatively intact subhaloes, to tidal streams originating from completely disrupted subhaloes as well as combinations of the two. A substructure identified by \STF\ can be any one of these substructures. However, by examining the bulk properties of a substructure it should be possible to identify its specific type. For instance, streams are likely to have anisotropic velocity distributions, whereas intact subhaloes likely have isotropic distributions. Subhaloes are physical overdensities, whereas streams are diffuse and correspond to a small fraction of the mass in the volume occupied by the stream. We use three parameters to classify a group; the overdensity, compactness and velocity isotropy. The overdensity of a group is determined by the the average physical density contrast between the group and the background, $\logrho$, which is calculated in a similar fashion as $\logphase$ limited to configuration space. The compactness is given by the volume mass fraction, 
$\volfrac$, that is the fraction of mass belonging to the group in the volume occupied by the group. The velocity isotropy, $\veliso$, is given by
\begin{align}
    \veliso=\frac{\sigma_2+\sigma_3}{2\sigma_1},
\end{align}
where $\sigma_i$ are the ordered eigenvalues of the group's velocity dispersion tensor $\sigma^2_{i,j}$. If the dispersion is isotropic $\veliso\sim1$. 
\begin{figure}
    \centering
    \includegraphics[width=0.45\textwidth]{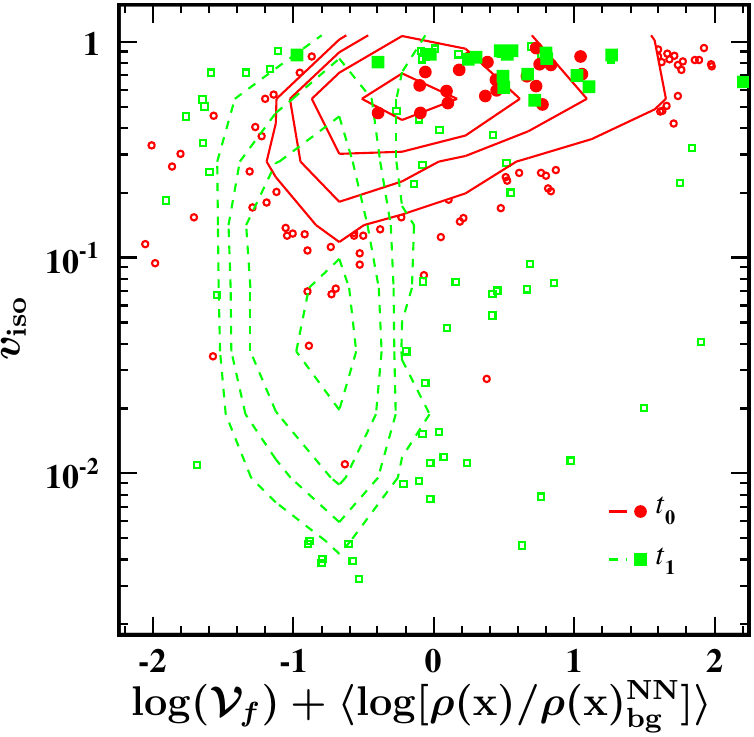}
    \caption{The distribution of groups found by \STF-3.0 in the  $\log(\volfrac)+\logrho$-$\veliso$ plane at two different times. We plot contours corresponding enclosing $0.75$, $0.5$, $0.25$ and $0.1$ of the population along with outliers (open points). We also show a sample of groups that correspond to bound subhaloes (large filled points).}
    \label{fig:halomany-vfrhovelaniso}
\end{figure}

\par
We plot the groups identified by \STF-3.0 in the $\log(\volfrac)+\logrho$-$\veliso$ plane in \Figref{fig:halomany-vfrhovelaniso}. For clarity, we plot contours of the groups found at $t_0$ and $t_1$ and outliers of these contours. We also check to see if a group is self-bound and plot a sample of bound groups for clarity since they are clustered in one region of this plane. Initially, most groups are compact overdensities, with $\log(\volfrac)+\logrho\gtrsim0$, and have isotropic velocities. Most of these groups are self-bound subhaloes, accounting for $\approx25\%$ of the halo's mass. The few groups with very anisotropic velocities dispersions which are neither overdense nor compact are unbound substructures. By $t_1$, dynamical processes have disrupted these subhaloes, causing some to form tidal tails and others to be completely disrupted. A substantial fraction of the groups, $\approx70\%$, at this time have $\veliso\lesssim10^{-1}$ and $\log(\volfrac)+\logrho\lesssim-0.5$. Only $\approx4\%$ of the groups identified at $t_1$ are subhalo like, that is groups with $\log(\volfrac)+\logrho\gtrsim0$ and $\veliso\sim1$. The bound subhaloes identified at $t_1$, (large filled square points), account for $\approx2\%$ of the halo's mass. The few groups identified with anisotropic velocities but $\log(\volfrac)+\logrho\gtrsim0$ are extended tidal tails with the core of the subhalo embedded inside. The remainder are unbound tidal streams. By $t_2$, the fraction of groups we would classify as subhaloes has decreased to $\approx2\%$ while the fraction of tidal stream groups has remained relatively constant.

\subsection{Inner Region}
\begin{figure*}
    \centering
    \includegraphics[width=0.425\textwidth]{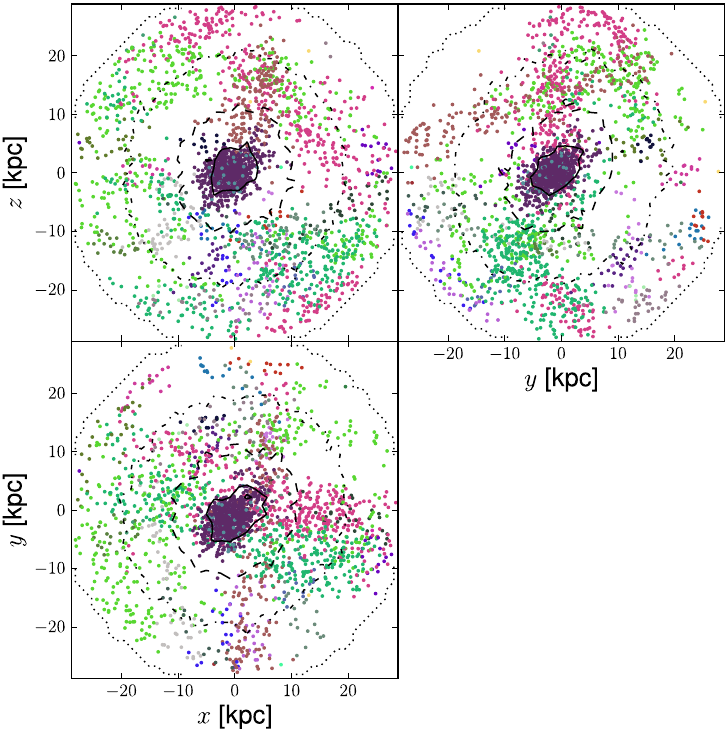}
    \includegraphics[width=0.425\textwidth]{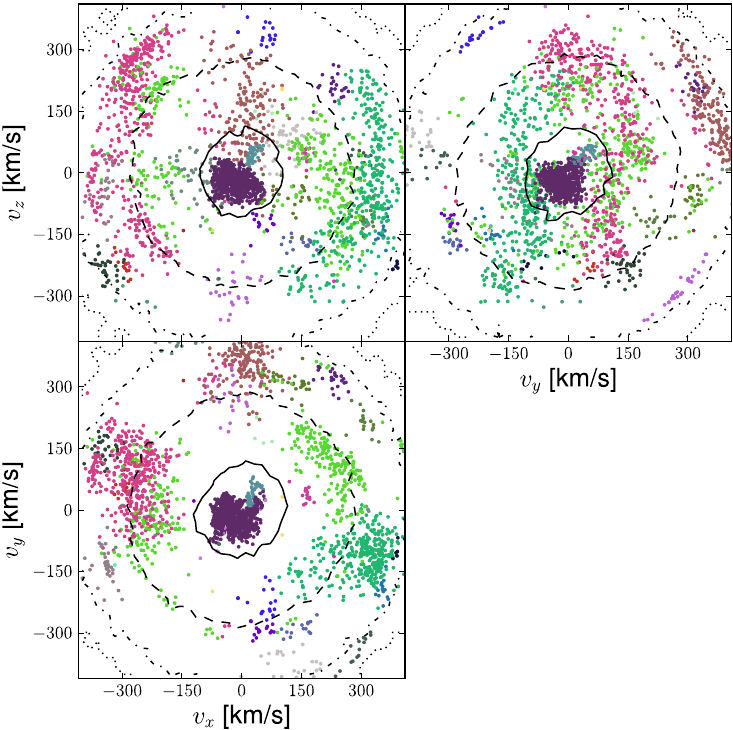}
    \caption{Positions (left) \& velocities (right) of particles belonging to groups identified by \STF-IT in the central $r\leq2a_{o,{\rm H}}$ region of halo at $t_2$. Particles are colour coded according to their group. The contours in the right panel enclose the projected positions of $25\%$, $50\%$, \& $75\%$ of all the particles in the central region. The velocity contours correspond to $1\sigma$, $2\sigma$ and $3\sigma$ levels enclosing $68\%$, $95\%$, $99\%$ of the all particles in this central region.}
    \label{fig:halomany-center}
\end{figure*}
As previously mentioned, \Figref{fig:halomany-massfunc} suggests that there are numerous substructures located in the very inner region of the halo at late times. However, due to the complicated morphology and extended nature of tidal streams, the center-of-mass need not even be located in the substructure itself. As a result this radial distribution may be misleading. In order to verify if we can find substructures in the very central regions of a halo, we plot the positions and velocities of all particles belonging to an \STF\ group at $t_2$ in \Figref{fig:halomany-center}. This figure clearly shows that we do indeed find numerous streams crossing this region.

\par
These velocity substructures account for $\sim1\%$ of the mass in this central region. Some of these particles belong to large tidal streams that were produced by massive subhaloes which lie outside this region. Other particles belong to small substructures that are located in this region. An example of such a substructure is the spheroidal velocity substructure located at the center. The group has a high purity, $p\approx0.6$, and primarily originates from a single subhalo at $t_0$. Most of particles identified have high relative velocities, lying outside the background's $68\%$ contour. Roughly $70\%$ of the particles have velocities that are $\gtrsim2\sigma$ away from the mean velocity. Notably, not all the groups identified have high velocities. The spherical substructure located near the center is actually within the velocity distribution of the background but is dynamically much colder.

\par
This figure indicates that that the central regions of haloes are not devoid of substructure even if the phase-space structure appears relatively smooth. It also clearly demonstrates that \STF\ is capable of identifying these structures even in a low resolution N-body realizations of haloes containing $\sim10^{6}$ particles.

\section{Discussion and Conclusion}\label{sec:discussion}
In this paper, we present an algorithm designed to identify both tidal streams and subhaloes.  We begin by dividing the halo into cells. In each cell, we model the velocity distribution function of the background as a multivariate Gaussian (the ``Maxwellian sea'').  While there is no exact method for identifying and modelling the background population, we have demonstrated that decomposing the halo into sub-volumes (cells) containing a relatively large number of particles is sufficient to average out any substructures present. We then calculate an effective velocity-space density for each particle in the cell. The ratio of the {\em local} velocity density to the density of the background at the particle's velocity provides a probabilistic measure of whether the particle belongs to the background or to a stream or subhalo. This key step of identifying outliers is then combined with a FOF-like search to identify substructures. 

\par
We have clearly shown using several test cases that our method is capable of identifying tidal streams and any subhaloes present in a halo. The purity of the identified substructures is very high regardless of the type of substructure, whether subhalo or diffuse tidal stream. Our method is even capable of identifying streams with a high purity in cases where the particles of the stream do not appear significantly overdense in either physical or phase-space for a given phase-space estimator, and account for a small fraction of the mass in the spatial volume occupied by the stream. Furthermore, we are able to effectively identify substructures in the cores of haloes, even haloes with low resolution.

\par
We also note that our key step of identifying velocity outliers can also be applied to any quantity for which one can characterize the mean distribution. The increased contrast between particles belonging to substructures and those belonging to the  background makes searching the resulting subset simpler. Furthermore, one could combine this key step with a variety of searches tailored to identify specific substructures. 

\par
Our algorithm is not the only one capable of finding tidal streams. Both \hsf\ \citep{hsf} and \enlink\ \citep{enlink} should be able to find tidal streams, however, both these algorithms rely on accurately estimating the phase-space density for each particle. In general, determining the velocity density is less computationally intensive than accurately computing the phase-space density. Searching the phase-space of a halo requires methods that calculate the local six dimensional metric for each particle to accurately determine distances in phase-space. We have also shown that, in certain cases, identifying substructures solely based on the phase-space density may miss diffuse tidal streams.

\par
The presence of tidal streams in the central regions of haloes has potentially important ramifications for indirect and direct dark matter searches. For instance, most studies of direct dark matter detection assume the velocity distribution of dark matter in the solar neighbourhood of a galactic mass halo is Maxwellian in order to determine the detection rate and recoil spectrum. Simulations show that the velocity distribution is more complex, possibly due to the presence of substructure. A few studies have attempted to go beyond this simple assumption but are limited by the resolution of the studies used to determine the velocity distribution \citep{fairbairn2009, kuhlen2010}. Since our algorithm allow us to accurately characterize the stream distribution function with a single snapshot from a cosmological simulation, it provides a easily accessible avenue for going beyond the resolution of current simulations. By searching numerous haloes from numerous cosmological simulations and stacking results, we will be able to determine the distribution function of these streams and extrapolate it to improve the accuracy of detection rates and the observed energy spectrum.

\section*{Acknowledgements}
We thank Joachim Stadel for useful discussions during the early stages of the project. LMW wishes to thank the ITP at the University of Zurich for their hospitality during a recent sabatical visit. The authors thank the referee, Yago Ascasibar, for useful comments. PJE acknowledges financial support from the Chinese Academy of Sciences (CAS). RJT and LMW acknowledge funding by respective Discovery Grants from NSERC. RJT is also supported by grants from the Canada Foundation for Innovation and the Canada Research Chairs Program. Simulations and analysis were performed on the computing facilities at the {\em Computational Astrophysics Laboratory} at Saint Mary's University. 

%-------------------------
\bibliographystyle{mn2e}
%\bibliography{ref.bib}
\bibliography{stfofcode.bbl}

\begin{thebibliography}{}

\bibitem[\protect\citeauthoryear{Appel}{Appel}{1985}]{appel1985}
Appel A.,  1985, SIAM J. Sci. Stat. Comput., 6, 85

\bibitem[\protect\citeauthoryear{{Ascasibar} \& {Binney}}{{Ascasibar} \&
  {Binney}}{2005}]{fiestas}
{Ascasibar} Y.,  {Binney} J.,  2005, \mnras, 356, 872

\bibitem[\protect\citeauthoryear{{Aubert}, {Pichon} \& {Colombi}}{{Aubert}
  et~al.}{2004}]{aubert2004}
{Aubert} D.,  {Pichon} C.,    {Colombi} S.,  2004, \mnras, 352, 376

\bibitem[\protect\citeauthoryear{{Barnes} \& {Hut}}{{Barnes} \&
  {Hut}}{1986}]{treecode}
{Barnes} J.,  {Hut} P.,  1986, \nat, 324, 446

\bibitem[\protect\citeauthoryear{{Diemand}, {Kuhlen} \& {Madau}}{{Diemand}
  et~al.}{2006}]{diemand2006}
{Diemand} J.,  {Kuhlen} M.,    {Madau} P.,  2006, \apj, 649, 1

\bibitem[\protect\citeauthoryear{{Diemand}, {Kuhlen}, {Madau}, {Zemp}, {Moore},
  {Potter} \& {Stadel}}{{Diemand} et~al.}{2008}]{diemand2008}
{Diemand} J.,  {Kuhlen} M.,  {Madau} P.,  {Zemp} M.,  {Moore} B.,  {Potter} D.,
     {Stadel} J.,  2008, \nat, 454, 735

\bibitem[\protect\citeauthoryear{{Elahi}, {Thacker}, {Widrow} \&
  {Scannapieco}}{{Elahi} et~al.}{2009}]{elahi2009}
{Elahi} P.~J.,  {Thacker} R.~J.,  {Widrow} L.~M.,    {Scannapieco} E.,  2009,
  \mnras, 395, 1950

\bibitem[\protect\citeauthoryear{{Fairbairn} \& {Schwetz}}{{Fairbairn} \&
  {Schwetz}}{2009}]{fairbairn2009}
{Fairbairn} M.,  {Schwetz} T.,  2009, Journal of Cosmology and Astro-Particle
  Physics, 1, 37

\bibitem[\protect\citeauthoryear{Friedman, Bentley \& Finkel}{Friedman
  et~al.}{1977}]{kdtree}
Friedman J.~H.,  Bentley J.~L.,    Finkel R.~A.,  1977, ACM Trans. Math.
  Softw., 3, 209

\bibitem[\protect\citeauthoryear{{Gao}, {White}, {Jenkins}, {Stoehr} \&
  {Springel}}{{Gao} et~al.}{2004}]{gao2004}
{Gao} L.,  {White} S.~D.~M.,  {Jenkins} A.,  {Stoehr} F.,    {Springel} V.,
  2004, \mnras, 355, 819

\bibitem[\protect\citeauthoryear{{G{\'o}mez} \& {Helmi}}{{G{\'o}mez} \&
  {Helmi}}{2010}]{gomez2010}
{G{\'o}mez} F.~A.,  {Helmi} A.,  2010, \mnras, 401, 2285

\bibitem[\protect\citeauthoryear{{Helmi} \& {de Zeeuw}}{{Helmi} \& {de
  Zeeuw}}{2000}]{helmi2000}
{Helmi} A.,  {de Zeeuw} P.~T.,  2000, \mnras, 319, 657

\bibitem[\protect\citeauthoryear{{Helmi}, {White} \& {Springel}}{{Helmi}
  et~al.}{2003}]{helmi2003}
{Helmi} A.,  {White} S.~D.~M.,    {Springel} V.,  2003, \mnras, 339, 834

\bibitem[\protect\citeauthoryear{{Ibata}, {Lewis}, {Irwin}, {Totten} \&
  {Quinn}}{{Ibata} et~al.}{2001}]{ibata2001}
{Ibata} R.,  {Lewis} G.~F.,  {Irwin} M.,  {Totten} E.,    {Quinn} T.,  2001,
  \apj, 551, 294

\bibitem[\protect\citeauthoryear{{Ibata}, {Gilmore} \& {Irwin}}{{Ibata}
  et~al.}{1994}]{ibata1994}
{Ibata} R.~A.,  {Gilmore} G.,    {Irwin} M.~J.,  1994, \nat, 370, 194

\bibitem[\protect\citeauthoryear{{Johnston}, {Bullock}, {Sharma}, {Font},
  {Robertson} \& {Leitner}}{{Johnston} et~al.}{2008}]{johnston2008}
{Johnston} K.~V.,  {Bullock} J.~S.,  {Sharma} S.,  {Font} A.,  {Robertson}
  B.~E.,    {Leitner} S.~N.,  2008, \apj, 689, 936

\bibitem[\protect\citeauthoryear{{Knebe}, {Knollmann}, {Muldrew}, {Pearce},
  {Aragon-Calvo}, {Ascasibar}, {Behroozi}, {Ceverino}, {Colombi}, {Diemand},
  {Dolag}, {Falck}, {Fasel}, {Gardner}, {Gottl{\"o}ber}, {Hsu}, {Iannuzzi},
  {Klypin}, {Luki{\'c}}, {Maciejewsk},  {McBride},  {Neyrinck},
  {Planelles},  {Potter},  {Quilis},  {Rasera},  {Read},
  {Ricker},  {Roy},  {Springel},  {Stadel},  {Stinson},
  {Sutter},  {Turchaninov},  {Tweed},  {Yepes}, {Zemp}}{{Knebe} et~al.}{2011}]{knebe2011}
  {Knebe} A.,  {Knollmann} S.~R.,  {Muldrew} S.~I.,  {Pearce} F.~R.,
  {Aragon-Calvo} M.~A.,  {Ascasibar} Y.,  {Behroozi} P.~S.,  {Ceverino} D.,
  {Colombi} S.,  {Diemand} J.,  {Dolag} K.,  {Falck} B.~L.,  {Fasel} P.,
  {Gardner} J.,  {Gottl{\"o}ber} S.,  {Hsu} C.-H.,  {Iannuzzi} F.,  {Klypin}
  A.,  {Luki{\'c}} Z.,  {Maciejewski} M.,  {McBride} C.,  {Neyrinck} M.~C.,
  {Planelles} S.,  {Potter} D.,  {Quilis} V.,  {Rasera} Y.,  {Read} J.~I.,
  {Ricker} P.~M.,  {Roy} F.,  {Springel} V.,  {Stadel} J.,  {Stinson} G.,
  {Sutter} P.~M.,  {Turchaninov} V.,  {Tweed} D.,  {Yepes} G.,    {Zemp} M.,
  2011, \mnras, pp 819--+

\bibitem[\protect\citeauthoryear{{Kuhlen}, {Weiner}, {Diemand}, {Madau},
  {Moore}, {Potter}, {Stadel} \& {Zemp}}{{Kuhlen} et~al.}{2010}]{kuhlen2010}
{Kuhlen} M.,  {Weiner} N.,  {Diemand} J.,  {Madau} P.,  {Moore} B.,  {Potter}
  D.,  {Stadel} J.,    {Zemp} M.,  2010, Journal of Cosmology and
  Astro-Particle Physics, 2, 30

\bibitem[\protect\citeauthoryear{{Kuijken} \& {Dubinski}}{{Kuijken} \&
  {Dubinski}}{1995}]{galactics1995}
{Kuijken} K.,  {Dubinski} J.,  1995, \mnras, 277, 1341

\bibitem[\protect\citeauthoryear{{Lewin} \& {Smith}}{{Lewin} \&
  {Smith}}{1996}]{lewin1996}
{Lewin} J.~D.,  {Smith} P.~F.,  1996, Astroparticle Physics, 6, 87

\bibitem[\protect\citeauthoryear{{Maciejewski}, {Colombi}, {Springel}, {Alard}
  \& {Bouchet}}{{Maciejewski} et~al.}{2009}]{hsf}
{Maciejewski} M.,  {Colombi} S.,  {Springel} V.,  {Alard} C.,    {Bouchet}
  F.~R.,  2009, \mnras, 396, 1329

\bibitem[\protect\citeauthoryear{{McConnachie}, {Irwin}, {Ibata}, {Dubinski},
  {Widrow}, {Martin}, {C{\^o}t{\'e}}, {Dotter}, {Navarro}, {Ferguson}, {Puzia},
  {Lewis}, {Babul}, {Barmby}, {Bienaym{\'e}}, {Chapman}, {Cockcroft},
  {Collins}, {Fardal}, {Harris}, {Huxor}, {Mackey}, {Pe{\~n}arrubia}, {Rich}, {Richer},
  {Siebert}, {Tanvir}, {Valls-Gabaud} \& {Venn}}{{McConnachie} et~al.}{2009}]{pandas2009}
  {McConnachie} A.~W.,  {Irwin} M.~J.,  {Ibata} R.~A.,  {Dubinski} J.,  {Widrow}
  L.~M.,  {Martin} N.~F.,  {C{\^o}t{\'e}} P.,  {Dotter} A.~L.,  {Navarro}
  J.~F.,  {Ferguson} A.~M.~N.,  {Puzia} T.~H.,  {Lewis} G.~F.,  {Babul} A.,
  {Barmby} P.,  {Bienaym{\'e}} O.,  {Chapman} S.~C.,  {Cockcroft} R.,
  {Collins} M.~L.~M.,  {Fardal} M.~A.,  {Harris} W.~E.,  {Huxor} A.,  {Mackey}
  A.~D.,  {Pe{\~n}arrubia} J.,  {Rich} R.~M.,  {Richer} H.~B.,  {Siebert} A.,
  {Tanvir} N.,  {Valls-Gabaud} D.,    {Venn} K.~A.,  2009, \nat, 461, 66

\bibitem[\protect\citeauthoryear{{Mignard}, {Bailer-Jones}, {Bastian},
  {Drimmel}, {Eyer}, {Katz}, {van Leeuwen}, {Luri}, {O'Mullane}, {Passot},
  {Pourbaix} \& {Prusti}}{{Mignard} et~al.}{2008}]{gaia2008}
{Mignard} F.,  {Bailer-Jones} C.,  {Bastian} U.,  {Drimmel} R.,  {Eyer} L.,
  {Katz} D.,  {van Leeuwen} F.,  {Luri} X.,  {O'Mullane} W.,  {Passot} X.,
  {Pourbaix} D.,    {Prusti} T.,  2008, in {W.~J.~Jin, I.~Platais, \&
  M.~A.~C.~Perryman} ed., IAU Symposium Vol.~248 of IAU Symposium, {Gaia:
  organisation and challenges for the data processing}.
pp 224--230

\bibitem[\protect\citeauthoryear{{Moore}, {Ghigna}, {Governato}, {Lake},
  {Quinn}, {Stadel} \& {Tozzi}}{{Moore} et~al.}{1999}]{moore1999}
{Moore} B.,  {Ghigna} S.,  {Governato} F.,  {Lake} G.,  {Quinn} T.,  {Stadel}
  J.,    {Tozzi} P.,  1999, \apjl, 524, L19

\bibitem[\protect\citeauthoryear{{Navarro}, {Frenk} \& {White}}{{Navarro}
  et~al.}{1997}]{nfw}
{Navarro} J.~F.,  {Frenk} C.~S.,    {White} S.~D.~M.,  1997, \apj, 490, 493

\bibitem[\protect\citeauthoryear{{Sharma} \& {Johnston}}{{Sharma} \&
  {Johnston}}{2009}]{enlink}
{Sharma} S.,  {Johnston} K.~V.,  2009, \apj, 703, 1061

\bibitem[\protect\citeauthoryear{{Sharma} \& {Steinmetz}}{{Sharma} \&
  {Steinmetz}}{2006}]{enbid}
{Sharma} S.,  {Steinmetz} M.,  2006, \mnras, 373, 1293

\bibitem[\protect\citeauthoryear{{Springel}}{{Springel}}{2005}]{gadget2}
{Springel} V.,  2005, \mnras, 364, 1105

\bibitem[\protect\citeauthoryear{{Springel}, {Wang}, {Vogelsberger}, {Ludlow},
  {Jenkins}, {Helmi}, {Navarro}, {Frenk} \& {White}}{{Springel}
  et~al.}{2008}]{springel2008}
{Springel} V.,  {Wang} J.,  {Vogelsberger} M.,  {Ludlow} A.,  {Jenkins} A.,
  {Helmi} A.,  {Navarro} J.~F.,  {Frenk} C.~S.,    {White} S.~D.~M.,  2008,
  \mnras, 391, 1685

\bibitem[\protect\citeauthoryear{{Springel}, {White}, {Tormen} \&
  {Kauffmann}}{{Springel} et~al.}{2001}]{subfind}
{Springel} V.,  {White} S.~D.~M.,  {Tormen} G.,    {Kauffmann} G.,  2001,
  \mnras, 328, 726

\bibitem[\protect\citeauthoryear{{Stadel}}{{Stadel}}{2001}]{skid}
{Stadel} J.~G.,  2001, PhD thesis, AA(UNIVERSITY OF WASHINGTON)

\bibitem[\protect\citeauthoryear{{Stiff} \& {Widrow}}{{Stiff} \&
  {Widrow}}{2003}]{stiff2003}
{Stiff} D.,  {Widrow} L.~M.,  2003, Physical Review Letters, 90, 211301

\bibitem[\protect\citeauthoryear{{Stiff}, {Widrow} \& {Frieman}}{{Stiff}
  et~al.}{2001}]{stiff2001}
{Stiff} D.,  {Widrow} L.~M.,    {Frieman} J.,  2001, \prd, 64, 083516

\bibitem[\protect\citeauthoryear{{Vogelsberger}, {Helmi}, {Springel}, {White},
  {Wang}, {Frenk}, {Jenkins}, {Ludlow} \& {Navarro}}{{Vogelsberger}
  et~al.}{2009}]{vogelsberger2009b}
{Vogelsberger} M.,  {Helmi} A.,  {Springel} V.,  {White} S.~D.~M.,  {Wang} J.,
  {Frenk} C.~S.,  {Jenkins} A.,  {Ludlow} A.,    {Navarro} J.~F.,  2009,
  \mnras, 395, 797

\bibitem[\protect\citeauthoryear{{Vogelsberger} \& {White}}{{Vogelsberger} \&
  {White}}{2010}]{vogelsberger2010}
{Vogelsberger} M.,  {White} S.~D.~M.,  2010, ArXiv e-prints

\bibitem[\protect\citeauthoryear{{Vogelsberger}, {White}, {Helmi} \&
  {Springel}}{{Vogelsberger} et~al.}{2008}]{vogelsberger2008}
{Vogelsberger} M.,  {White} S.~D.~M.,  {Helmi} A.,    {Springel} V.,  2008,
  \mnras, 385, 236

\bibitem[\protect\citeauthoryear{{Vogelsberger}, {White}, {Mohayaee} \&
  {Springel}}{{Vogelsberger} et~al.}{2009}]{vogelsberger2009}
{Vogelsberger} M.,  {White} S.~D.~M.,  {Mohayaee} R.,    {Springel} V.,  2009,
  ArXiv e-prints

\bibitem[\protect\citeauthoryear{{Widrow} \& {Dubinski}}{{Widrow} \&
  {Dubinski}}{2005}]{galactics}
{Widrow} L.~M.,  {Dubinski} J.,  2005, \apj, 631, 838

\bibitem[\protect\citeauthoryear{{Widrow}, {Pym} \& {Dubinski}}{{Widrow}
  et~al.}{2008}]{widrow2008}
{Widrow} L.~M.,  {Pym} B.,    {Dubinski} J.,  2008, \apj, 679, 1239

\bibitem[\protect\citeauthoryear{{Xu}, {Mao}, {Wang}, {Springel}, {Gao},
  {White}, {Frenk}, {Jenkins}, {Li} \& {Navarro}}{{Xu} et~al.}{2009}]{xu2009}
{Xu} D.~D.,  {Mao} S.,  {Wang} J.,  {Springel} V.,  {Gao} L.,  {White}
  S.~D.~M.,  {Frenk} C.~S.,  {Jenkins} A.,  {Li} G.,    {Navarro} J.~F.,  2009,
  \mnras, 398, 1235

\end{thebibliography}

\end{document}